\documentclass[useAMS,usenatbib]{mnras}
\usepackage[utf8x]{inputenc}

\usepackage{graphicx}
\usepackage[T1]{fontenc}
\usepackage{aecompl}
\usepackage[dvipsnames]{color}
\usepackage[normalem]{ulem}
\usepackage{threeparttable}
\usepackage{multirow}
\usepackage{array}
\usepackage{journals}

\newcommand{\SII}{[S~{\sc ii}]\ }
\newcommand{\OIII}{[O~{\sc iii}]\ }
\newcommand{\NII}{[N~{\sc ii}]\ }
\newcommand{\HII}{H~{\sc ii}\ }

\newcommand{\HI}{H~{\sc i}\ }
\newcommand{\Ha}{H$\alpha$\ }
\newcommand{\kms}{\,\mbox{km}\,\mbox{s}^{-1}}

\newcommand{\HST}{\textit{HST}\ }
\newcommand{\SIIHa}{[S~{\sc ii}]/H$\alpha$}
\newcommand{\NIIHa}{[N~{\sc ii}]/H$\alpha$}
\newcommand{\OIIIHb}{[O~{\sc iii}]/H$\beta$}
\newcommand{\be}{\begin{equation}}
\newcommand{\ee}{\end{equation}}

\newcommand{\revone}{}

\def \gtsima{$\, \buildrel > \over \sim \,$}
\def \ltsima{$\, \buildrel < \over \sim \,$}
\def \simgt{\lower.5ex\hbox{\gtsima}}
\def \simlt{\lower.5ex\hbox{\ltsima}}

\usepackage{etoolbox}
\makeatletter
\newcount\c@additionalboxlevel
\newcount\c@maxboxlevel
\title[Star formation in the galaxy DDO~53.] {Star formation in the nearby dwarf galaxy DDO~53:  interplay between gas accretion and stellar feedback.}

\author[Egorov et al.]{
	Oleg~V.~Egorov$^{1,2}$\thanks{E-mail: oleg.egorov@uni-heidelberg.de},
	Tatiana~A.~Lozinskaya$^{2}$,
	Konstantin~I.~Vasiliev$^{3}$,	\newauthor
	Anastasiya~D.~Yarovova$^{2,3}$,
	Ivan~S.~Gerasimov$^{3}$,
	Kathryn Kreckel$^{1}$,
	Alexei~V.~Moiseev$^{4,2}$
	\\
	$^{1}$ Astronomisches Rechen-Institut, Zentrum f\"{u}r Astronomie der Universit\"{a}t Heidelberg, M\"{o}nchhofstra\ss e 12-14, 69120 Heidelberg, Germany
	\\
	$^{2}$ Lomonosov Moscow State University, Sternberg Astronomical Institute,
	Universitetsky pr. 13, Moscow 119234, Russia
	\\
	$^{3}$ Lomonosov Moscow State University, Faculty of Physics,
	Leninskie gory 1-2, Moscow 119991, Russia
	\\
	$^{4}$ Special Astrophysical Observatory, Russian Academy of Sciences, Nizhnii Arkhyz 369167, Russia
}

\date{Accepted 2021 Month 00. Received 2021 Month 00; in original form 2021 Month 00}

\pagerange{\pageref{firstpage}--\pageref{lastpage}} \pubyear{2021}

\begin{document}

\maketitle

\label{firstpage}

\begin{abstract}
	We present the results of a multiwavelength study of the nearby dwarf galaxy DDO~53 -- a relatively isolated member of the M~81 group. 
	 We analyse the atomic and ionised gas kinematics (based on the observations with Fabry-Perot interferometer in \Ha line and archival data in \HI 21 cm line), distribution, excitation and oxygen abundance of the ionised gas (based on the long-slit and integral-field spectroscopy and on imaging with narrow-band filters), and their relation with the young massive stars (based on archival \HST data). We detect a faint 2-kpc sized supershell of ionised gas surrounding the galaxy. Most probably, this structure represents a large-scale gas outflow, however it could be also created by the ionising quanta leaking from star-forming regions to the marginally detected atomic hydrogen surrounding the galactic disc. We analyse the properties of the anomalous \HI in the north part of the galaxy and find that its peculiar kinematics is also traced by ionised gas. We argue that this \HI feature is related to the accreting gas cloud captured from the intergalactic medium or remaining after the merger event occurred $>1$~Gyr ago. The infalling gas produces shocks in the interstellar medium and could support the star formation activity in the brightest region in DDO~53.

\end{abstract}


\section{Introduction.}

Feedback from massive stars has far-reaching effects on a galaxy's interstellar medium (ISM) and is thought to play a crucial role in the life cycle of the molecular material and the regulation of star formation on global scales. 
Through three main channels (ionising radiation, stellar winds and supernovae explosions) massive stars carve out low density bubbles and shells (see, e.g., \citealt{Krumholz2014, Rahner2019}). These processes are especially important in dwarf irregular (dIrr) galaxies, where their thick gas-rich discs and a lack of spiral density waves allow the development of  prominent structures in the ISM formed as a result of stellar feedback.
  Moreover, thanks to their shallow gravitation potential, feedback in dIrr galaxies can easily launch galactic winds in comparison with more massive systems, and thus dwarf galaxies appear to be a major pollutants of the intergalactic medium (IGM) \citep[e.g.][]{Ferrara2000, Emerick2019}. 
Thus, nearby dIrr galaxies are ideal laboratories for studying the effects of massive stellar feedback on the surrounding material and could provide important constraints on the feedback parameters crucial for reproducing realistic galaxies in the cosmological simulations \citep[e.g.][]{Ceverino2009, Schaye2015, Keller2015}.

Beyond the Local Group, the galaxies from the nearest M81 group ($D\sim3.5$~Mpc) appear to be a perfect test-bed for studies of the feedback processes and the interactions of the galaxies with the IGM at the scales of individual \HII regions and massive stars. The group is reach in neutral hydrogen \citep{Sorgho2019} and thus provides a large diversity between the member galaxies in terms of their star formation rates (SFR), morphologies, masses, and metallicities \citep[e.g.][]{Karachentsev2002, Weisz2008, Croxall2009}. Based on the observations performed with high-resolution Fabry-Perot interferometer (FPI) and long-slit or integral field unit (IFU) spectrographs, we previously analysed in details the kinematics, morphology and chemical abundance of the ISM and their relation with the ongoing star formation activity in several individual galaxies of the M81 group: VII~Zw~403 \citep{Lozinskaya2006, Arkhipova2007, Egorov2011}, IC~2574 \citep{Egorov2014}, Holmberg~I \citep{Egorov2018}, Holmberg~II \citep{Egorov2013, Wiebe2014, Egorov2017}, NGC~3077 \citep{Oparin2020}. All these galaxies are gas rich and reveal extended complexes of star formation with prominent feedback-driven superbubbles, and in most of them the long duration of star formation activity led to the creation of so called supergiant shells (SGS), having sizes up to 2 kpc, the interaction of which could be a driver of star formation propagation \citep{Egorov2017, Vasiliev2020}.  \HI observations revealed a presence of gaseous tidal tails or surrounding high velocity clouds around some of the mentioned galaxies thus indicating their interaction with the IGM in the M81 group \citep{Ashley2017, Sorgho2019, Sorgho2020}.

Here we analyse the ISM and ongoing star formation in the quiescent and metal-poor dIrr galaxy DDO~53, which is located at the periphery of the M81 group at a distance of 3.68~Mpc (the general parameters derived in this paper or adopted from previous studies are given in Table~\ref{tab:ddo53}). In contrast with the previously mentioned galaxies, DDO~53 doesn't exhibit SGS in its ISM, or clear signs of tidal/accretion tails or surrounding gas clouds in its \HI maps. The nearest galaxy to DDO~53 -- blue compact dwarf galaxy UGC~4483 -- is located at $\sim223$ kpc from it \citep{Karachentsev2002}, that makes DDO~53 a fairly isolated galaxy. Despite that, several observational findings have drawn our attention to this galaxy and motivated us to perform a detailed analysis of its ISM: peculiar kinematics in the \HI at the periphery of the galactic disc and the presence of extended \Ha emission surrounding the galaxy (first reported in this paper).

Modelling the kinematics of atomic hydrogen, \citet{Iorio2017} found extra emission in the \HI 21 cm line in the north part the galaxy and suggested that it could be related to a gas outflow or inflow. In their recent work \cite{Hunter2019} have found only one region of strong non-circular motions of \HI in DDO~53, which is also located there, and one of the 7 \HI holes detected in this galaxy by \cite{Pokhrel2020} also coincides with this region.
From earlier  analysis of the galaxy's environment \citet{Pustilnik2003} suggested that the current starburst in DDO~53 could be triggered by the tidal disturbance by the M~81 group as whole, or by the interaction with the IGM. On the other hand, \cite{Begum2006} showed that the IGM ram pressure is unlikely to influence the gas distribution and kinematics in the galaxy because the estimated gas density in the vicinity of DDO~53 is much lower than demanded to explain the observed peculiarities. 
The authors speculated that DDO~53 may be a product of a recent merger between two faint dwarf galaxies. However, as was found by \cite{Weisz2008} from the analysis of the resolved star formation history, such a merger event should have occurred more than 1 Gyr ago, that means it is unlikely to be responsible for the observed non-circular motions and peculiarities in the \HI morphology, or for the recent burst of star formation that occurred about 25~Myr ago. The question on the origin of ongoing star-formation and of non-circular motions remains unsolved.

The optical appearance of DDO~53 is dominated by a few bright blue clumps, which emit copious amounts of H$\alpha$ and are observed towards the two high-density peaks of the \HI distribution \citep{Begum2006}. The kinematics of ionised gas of DDO~53 was first investigated with a scanning Fabry--Perot interferometer (FPI) by \citet{Dicaire2008}, but the authors were only able to measure the distribution of radial velocities in the bright \HII regions. \citet{Moiseev2012}  presented much deeper observations of this galaxy with FPI and  identified three areas of elevated velocity dispersion in DDO~53. Among them, only one is observed towards the bright \HII regions and demonstrates a shell-like morphology and thus could be easily interpreted as a feedback-driven superbubble. Two other structures are rather embedded in the diffuse ionised gas (DIG)  between the bright \HII regions, and since the information about the mechanism of gas excitation in these regions was unavailable, their origin remained uncertain. Beside these features in the ionised gas, we also discovered a faint 2-kpc sized \Ha shell-like structure surrounding the galaxy (we describe it further in the current paper). All these features point to the  far reaching impact of stellar feedback on the DDO~53 morphology.

In the current study we perform an in-depth analysis of the ISM of DDO~53 and its relation to the ongoing star formation and the surrounding IGM. We use new data from long-slit and integral-field spectroscopy and narrow-band imaging, as well as the the data obtained \revone{earlier} with a scanning FPI, 
 and the archival data of \HI observations from LITTLE THINGS survey \cite{Hunter2012}. The aim of the current analysis is to reveal the nature of the observed features in neutral and ionised gas morphology and kinematics mentioned above. The paper is organized as follows: Secton~\ref{sec:obs} contains the description of observational data; in Section~\ref{sec:results} we report the results of the observations. Namely, Sections~ \ref{sec:HII} describes the morphology of ionised gas in DDO~53; Section~\ref{sec:HI} is dedicated to the morphology and kinematics of atomic component; Section~\ref{sec:local_kin} presents our analysis of the small-scale kinematics of ionised gas; and in Section~\ref{sec:spectra_res} we focus on the analysis of the gas excitation and metallicity using the spectral data. In Section~\ref{sec:discussion} we discuss the results obtained considering the potential role of the gas accretion and stellar feedback in the creation of the observed peculiarities in DDO~53's appearance. Section~\ref{sec:summary} summarises our results.

\begin{table}
	\begin{center}
\caption{General parameters of the galaxy DDO~53}
\label{tab:ddo53}
\begin{tabular}{lr}
\hline
Parameters & Value \\
\hline
Names$^a$ & DDO~53, UGC~4459, VII~Zw238,  \\
			&             PGC~24050, VV~499 \\
Distance$^b$ & 3.68 Mpc \\
$M_B^a$ & $-13.49^m$ \\
Optical radius$^a$, $R_{25}$ & 24 arcsec = 425 pc \\
$\log\mathrm{(SFR)}^c$ & $-2.2...-2.3$ \\
$12 + \log\mathrm{(O/H)}^d$ & $7.67 \pm 0.04$\\
$M_\mathrm{HI}^e$ & $7.1\times10^7 M_\odot$ \\
$M_{\star}^f$ & $1.0\times10^7 M_\odot$ \\
Inclination$^g$,  $i$ & $37^\circ$ \\ 
$PA_\mathrm{kin}^g$ & $123^\circ$ \\
$V_\mathrm{rot}^g$ & $20 \kms$ \\
\hline
\end{tabular}
	\begin{tablenotes}
	\scriptsize
	\item $^a$ HyperLEDA Database \\ \citep[][\url{http://http://leda.univ-lyon1.fr/}]{Makarov2014}.
	\item $^b$  \cite{Tully2009, Karach2013}
	\item $^c$ \cite{Hunter2010, Sabbi2018}
	\item $^d$ This work
	\item $^e$ \cite{Hunter2012}
	\item $^f$ \cite{Weisz2011}
	\item $^g$ \cite{Iorio2017}

\end{tablenotes}
	\end{center}
\end{table}

\section{Observations and data reduction}\label{sec:obs}

\subsection{Optical FPI-observations}
\label{sec_obs_ifp}

\begin{table*}
	\caption{Log of observational data}
	\label{tab:obs_data}
	\begin{tabular}{llrllllll}
		\hline
		Data set       & Date of obs    & $\mathrm{T_{exp}}$, s & FOV                             & $''/px$             & $\theta$, $''$   & sp. range              & $\lambda_c$, \AA              & $\delta\lambda$ or FWHM, \AA        \\ \hline
		FPI  & 2009 Feb 26 & $36\times200$  & {$6.1'\times6.1'$} & {0.71} & 3.3              & {13~\AA\, around \Ha} &  {$-$}  &{0.8~($35\kms$) } \\
		LS PA=198 & 2019 Feb 11 & 8400 &  {$1\arcsec\times6.1\arcmin$} & {0.36} & 2.8 & {3600--7070} &{$-$}   & 4.8 \\
		LS PA=211 &  2013 Nov 12 & 5400 & {$1\arcsec\times6.1\arcmin$} & {0.36} & 1.5 & {3600--7070} &{$-$} & 4.8 \\
		LS PA=259 & 2013 Nov 12 & 7200 &  {$1\arcsec\times6.1\arcmin$} & {0.36}  & 1.0 & {3600--7070} &{$-$}   & 4.8 \\
		LS PA=323 & 2013 Nov 11 & 9600 &  {$1\arcsec\times6.1\arcmin$} & {0.36} & 1.7 & {3600--7070} &{$-$}   & 4.8 \\
		Image FN655 & 2013 Dec 04 & 3000 & {$6.1\arcmin\times6.1\arcmin$} & {0.36} & 1.8 & \Ha+\NII & 6559  & 97 \\
		Image FN655 & 2019 Dec 18 & 3060 & {$6.1\arcmin\times6.1\arcmin$} & {0.36} & 1.7 & \Ha+\NII & 6559  & 97 \\
		Image FN674 &  2013 Dec 04 & 3000 & {$6.1\arcmin\times6.1\arcmin$} & {0.36} & 1.8 & \SII 6717+6731 & 6733  & 60  \\
		Image FN641 & 2013 Dec 04 & 1500 & {$6.1\arcmin\times6.1\arcmin$} & {0.36}  & 1.8 & continuum & 6413  & 179  \\
		Image FN641 & 2019 Dec 18 & 1020 & {$6.1\arcmin\times6.1\arcmin$} & {0.36}  & 1.7 & continuum & 6413  & 179  \\
		Image FN712 & 2019 Dec 18 & 1020 & {$6.1\arcmin\times6.1\arcmin$} & {0.36}  & 1.7 & continuum & 7137  & 209  \\
		PPAK IFU & 2013 Dec 15 & $3\times400$ & {$1\arcmin\times1\arcmin$} & 1.00 & 2.5 & 3700-7000 &  {$-$} & 10 \\ 
		\hline
	\end{tabular}
\end{table*}

The observations were performed at the prime focus of the 6-m
telescope of Special Astrophysical Observatory of the Russian Academy of Sciences (SAO RAS) using a scanning FPI mounted inside the
SCORPIO  multi-mode focal reducer \citep{scorpio}. These data were previously analysed and described in \citet{Moiseev2012,Moiseev2014,MK2015}.  
The log of these observations and the parameters of other data sets are given in Table~\ref{tab:obs_data}, where $\mathrm{T_{exp}}$ is the exposure time, FOV -- the field of view,  $''/px$ -- pixel size on the final images, $\theta$ -- the final angular resolution, $\lambda_c$ -- the central wavelength of the used filters, and $\delta\lambda$ is the final spectral resolution.


The result of the FPI observations and data reduction was a large--scale data cube containing 36-channel spectra in the region around the red-shifted  \Ha line. The analysis of the emission line profiles
was carried out using multi-component Voigt fitting \citep{Moiseev2008}, that yields flux, line-of-sight velocity and velocity dispersion (corrected for instrumental broadening) for each component as an output. The measured values of the \Ha velocity dispersion were additionally corrected for natural and thermal broadening by quadratically subtracting $\sigma_{th}=9.6\ \kms$.  

\subsection{Long-slit spectroscopic observations}
\label{sec_obs_ls}

The long slit observations were carried out with the SCORPIO-2 multi-mode focal reducer \citep{scorpio2} at the 6-m telescope of SAO RAS. Four spectra with different slit positions were obtained (Tab.~\ref{tab:obs_data}). 

The data reduction was performed in a standard way using the SCORPIO-2 \textsc{idl}-based  pipeline as described in \citet{Egorov2018}.  The observations of  the spectrophotometric standards BD+28d4211 at a close zenith distance immediately after or before the DDO~53 observations were  used  to calibrate its spectra to an absolute intensity scale.

To measure the fluxes of emission lines  our own \textsc{idl} software  based on the \textsc{mpfit} \citep{mpfit} routine was used. Gaussian fitting was applied to measure the integrated line fluxes of each studied region. For estimating the final uncertainties of the  line fluxes we quadratically added the errors propagated through all data reduction steps to the uncertainties returned by \textsc{mpfit}. We didn't performed any modelling or subtraction of the underlying stellar population because of its negligible contribution to the emission spectra.

A reddening correction was applied to each spectrum before estimating the lines flux ratios listed in this paper. For that we derived the colour excess $E(B-V)$ from the observed Balmer decrement and then used a \cite{Cardelli1989} curve parametrized by \cite{Fitzpatrick1999} to perform a reddening correction.
In this paper we use the following abbreviations for emission line flux ratios: \SIIHa\, is F([S~{\sc ii}] $\lambda6717+6731$\AA)/F(H$\alpha$); \NIIHa\, is F([N~{\sc ii}] $\lambda6584$\AA)/F(H$\alpha$); \OIIIHb\, is F([O~{\sc iii}] $5007$\AA)/F(H$\beta$).

\begin{figure}
	\includegraphics[width=\linewidth]{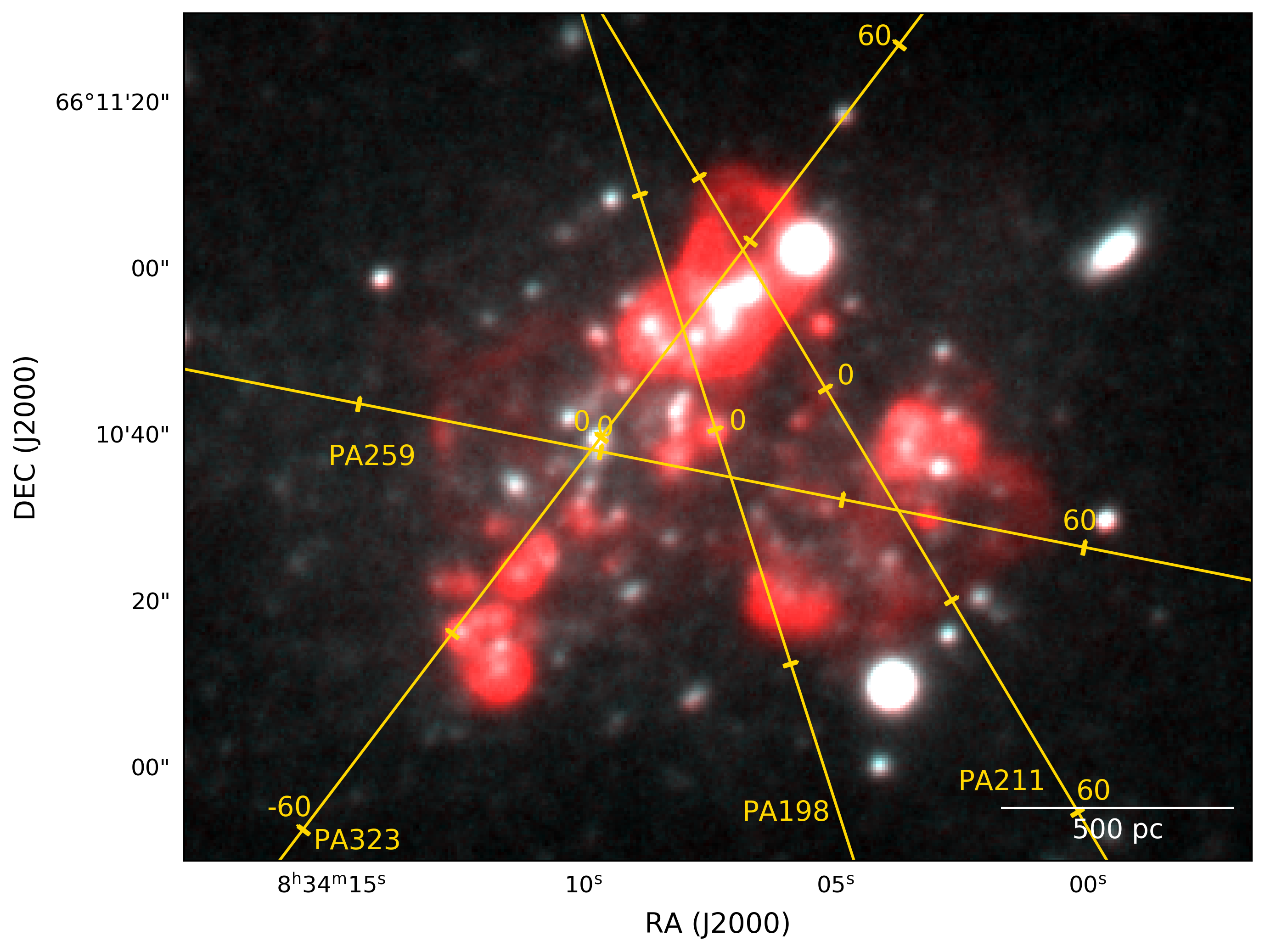}
	\caption{Position of the SCORPIO-2 slits  overlaid 
	on the image of DDO~53, where the red channel shows the distribution of H$\alpha$ emission (observed with FN655 filter) and the white colour traces the underlying continuum (in FN641 filter).
	}\label{fig:slitpos}
\end{figure}

\subsection{IFU spectroscopy }
The galaxy DDO~53 was observed with the PMAS spectrograph (Postdam Multi Aperture Spectrograph, \citealt{Roth2005}) in the PPaK mode \citep{Verheijen2004, Kelz2006} at the Calar Alto 3.5m telescope. This instrument samples a $\sim$1\arcmin\ hexagonal field of view with 331 fibers of 2.7\arcsec\ diameter. Observations were taken using the V300 grating, which provides full coverage of the optical spectral range (3700-7000\AA) while avoiding vignetting  from the instrument.  

Our data reduction follows the procedure described in detail in \citet{Kreckel2013}, and is summarized here. All data are reduced using the p3d package \citep{Sandin2010}, v2.2.5.1 `Serenity'. Due to the low $\sim$65 per cent filling factor of the fibers, the science field was observed in three dither positions. Arc and calibration lamp images are also obtained at the position of the science field in order to trace the dispersed fiber positions across the CCD.  Spectrophotometric standards were observed at the beginning and end of the night. Sky subtraction is carried out using simultaneously observed dedicated sky fibers placed $\sim$75\arcsec\ from the center of the field of view (see \citealt{Kelz2006}).  Emission line maps were constructed assuming a Gaussian line profile shape and fit using the same way as for long-slit data.

These IFU data provide the ability to trace the spatial distribution of the spectral properties of ionised gas, however we can perform such an analysis only for a few of the brightest lines (H$\alpha$, [O~\textsc{iii}], and  H$\beta$ to some extent), while the signal-to-noise (S/N) of the fainter lines (including, [N~\textsc{ii}], [S~\textsc{ii}]) is too small outside the brightest \HII regions. Since the long-slit spectra provide higher S/N, we utilize the IFU data only to analyse the spatial distribution of the \OIIIHb\ in the brightest star forming complex, and to measure the metallicities in a few bright regions not crossed by any of the slits. 

\subsection{Narrow-band imaging}
\label{sec_obs_img}

Deep optical images of DDO~53 in the H$\alpha$ and \SII emission lines were taken at the prime focus of the 6-m telescope of SAO RAS with SCORPIO-2 using narrow-band filters FN655 and FN674, respectively. The transmission curves of each filter can be found on the SCORPIO-2 web site\footnote{\url{https://www.sao.ru/hq/lsfvo/devices/scorpio-2/filters_eng.html}}.

To construct the final image in \Ha emission we combine the images obtained on different years at the same instrument and with the same set of filters. Each individual exposure is reduced separately. After standard data reduction processes including bias subtraction, flat-fielding, correction for variations of atmospheric extinction and seeing, all different exposures were aligned and combined using sigma-clipping to remove the cosmic hits and artefacts. 

We use the broader-band FN641 and FN712 filters centred on the continuum to subtract the stellar contamination from the obtained emission line images. The data reduction for the continuum images is performed in the same way as for the emission line images. The final images are aligned to the reference \Ha image using the \textsc{astroalign} procedure \citep{astroalign}. We combine both continuum images with the weights proportional to the distance of the central wavelength of the corresponding filter from that for the \Ha or \SII image. These final continuum images were subtracted from those obtained with the FN655 and FN674 filters to obtain our final pure \Ha and \SII emission line images. The \textsc{astrometry.net} service \citep{Lang2010} is used to do an astrometric calibration of the images. 

We also observed  spectrophotometric standards AGK+81d266 and BD+25d4655 during the same night in FN655 and FN674 filters, respectively, and use them for further flux calibration of our \Ha and \SII images. Our final narrow band images show surface brightness value at the $1\sigma$ level corresponding to $(1.8, 4.8) \times 10^{-18} \mathrm{erg\ s^{-1} cm^{-2} arcsec^{-2}}$ in \Ha and \SII emission lines, respectively, as measured by the standard deviation of the background intensities.


Note that because the FWHM of the FN655 filter is broader than the distance between the \Ha and [N~{\sc ii}]\, emission lines, the image in this filter is contaminated by [N~{\sc ii}] 6548, 6584~\AA\, emission. According to our spectral observations, the \NIIHa\, flux ratio is $0.03 - 0.06$ for DDO~53 (see Section~\ref{sec:spectra_res}). Because of the transmission of the FN655 filter is lower at the region of \NII emission lines than for H$\alpha$, we conclude that the \NII contamination of the \Ha images does not exceed 4 per cent.

\subsection{Other observational data used}
\label{sec_obs_other}

We use archival JVLA data of the \HI 21 cm line from the LITTLE THINGS survey \citep{Hunter2012} to study the \HI gas distribution and kinematics. In this work we analyse both the
natural-weighted (NA) and robust-weighted (RO) data cubes which have a velocity scale of $2.6 \kms$ per channel and  angular resolution of $beam_{NA}=11.8\times9.5$~arcsec and $beam_{RO}=6.3\times5.7$~arcsec, respectively.

The selection of the main sequence O-stars presented in this paper is based on the LEGUS catalogue \citep{Sabbi2018} providing photometry in 5 wide-band filters F275W, F336W, F438W, F555W and F814W (corresponding to the NUV, U, B, V, I bands) as observed with the \textit{Hubble} space telescope (\textit{HST}) using the ACS/WFC3 camera. We use the same criteria as in \citep{Kahre2018}, namely -- we limit the reddening-free parameter $Q=(m_{NUV}-m_{B}) - \frac{E(NUV-B)}{E(V-I)}(m_{V}-m_{I})$ to the range of  $-2.1\le Q\le0$. We also apply additional criteria following the model parameters for O stars from \cite{Martins2005}: $M_V<-3.9$ and $(m_V-m_I) \le 0.5$. Thus we select 85 O-star candidates which meet these criteria and consider further their distribution and properties. Their localisation is shown in Fig.~\ref{fig:regnames}.
Note that we use version 2.0 of the LEGUS catalogue, containing only stars with the highest-quality photometry, and hence the estimated number of O-stars is rather a lower limit. Using a less strict selection criteria would add about 40 additional O-stars candidates, but the general picture of their distribution remains unchanged. In addition, some stars towards the \HII regions might still be missed because of contamination of the measurements in the filters by line emission.


\section{Results of observations}\label{sec:results}
 

\subsection{Morphology of the \HII regions and diffuse ionised gas}\label{sec:HII}

\begin{figure}
	\includegraphics[width=\linewidth]{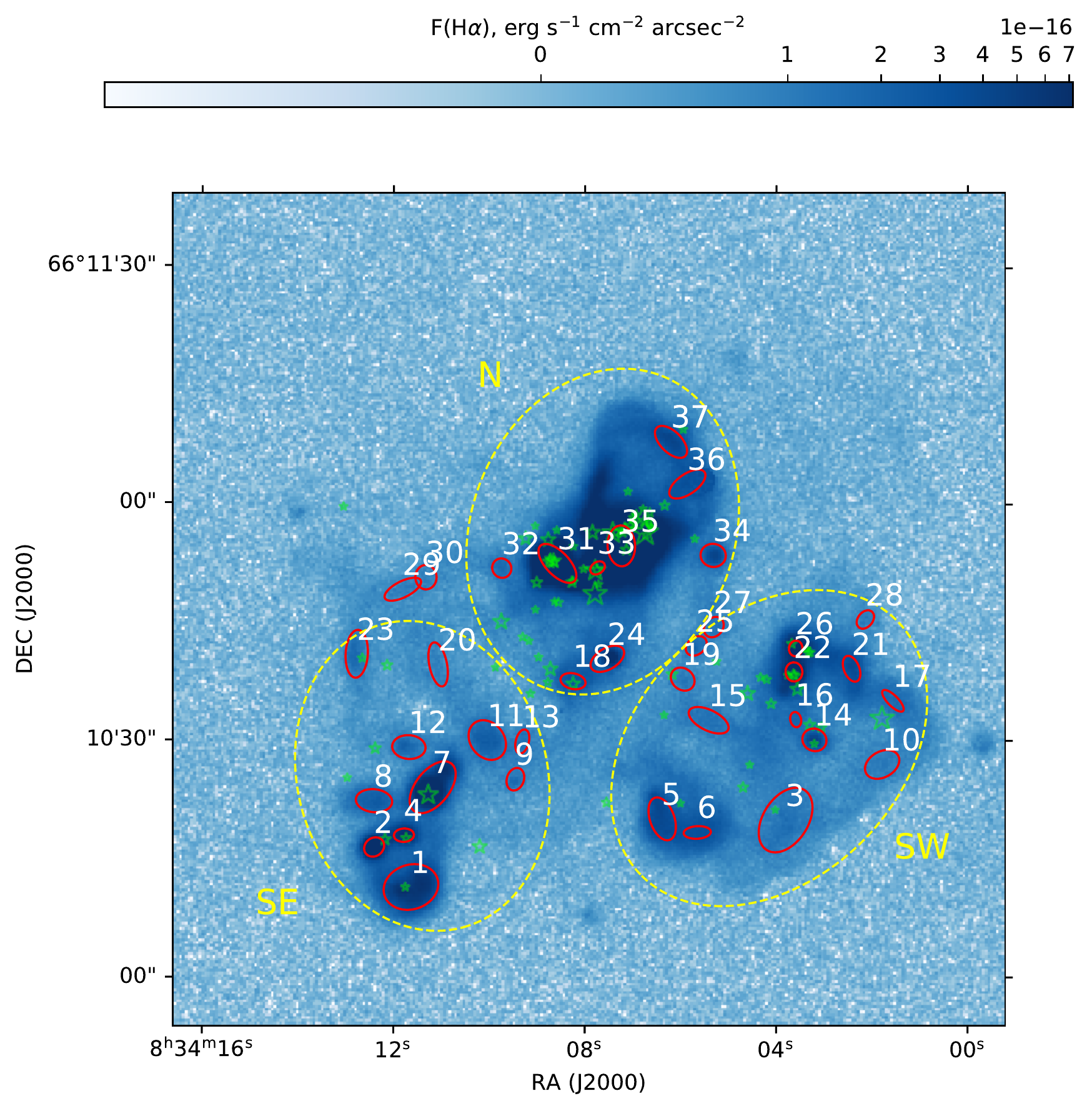}
	\caption{The location of \HII regions identified by the \textsc{astrodendro} routine overlaid on \Ha images. Yellow ellipses encircle the complexes of \HII regions according to our designation in the text. Pale green asterisks shows the location of the main sequence O stars identified in  the \HST  LEGUS catalogue; the size of the symbols correlate with $M_V$ of the star.}\label{fig:regnames}
\end{figure}

Optical images of DDO~53 reveal a few star-forming clumps emitting H$\alpha$ (see Fig.~\ref{fig:slitpos}, \ref{fig:regnames}).	The brightest northern \HII complex has a size of about 450 pc and consists of a number of resolved \HII regions and an extended ionised shell-like structure having a size of 270 pc. Other star-forming regions in the galaxy are remarkably fainter. A net of extended faint filamentary structures is observed in \Ha connecting the relatively bright \HII regions. 

\cite*{Strobel1990} analyzed their early images of DDO~53 in \Ha and identified 18 \HII regions in the galaxy. Since our images reveal  much fainter structures, we perform a new selection of the \HII regions using the  \textsc{astrodendro}\footnote{\url{http://www.dendrograms.org/}} routine, with  the following parameters:  the minimum level was set to be equal to 5$\sigma$ on our images, the minimal intensity difference between regions was set to 1$\sigma$, and the minimal size of the regions was set to the seeing value. We identified 37 \HII regions as shown in Fig.~\ref{fig:regnames}. All these regions could be roughly combined into 3 larger complexes denoted further as N, SE and SW. 

\begin{figure}
	\includegraphics[width=\linewidth]{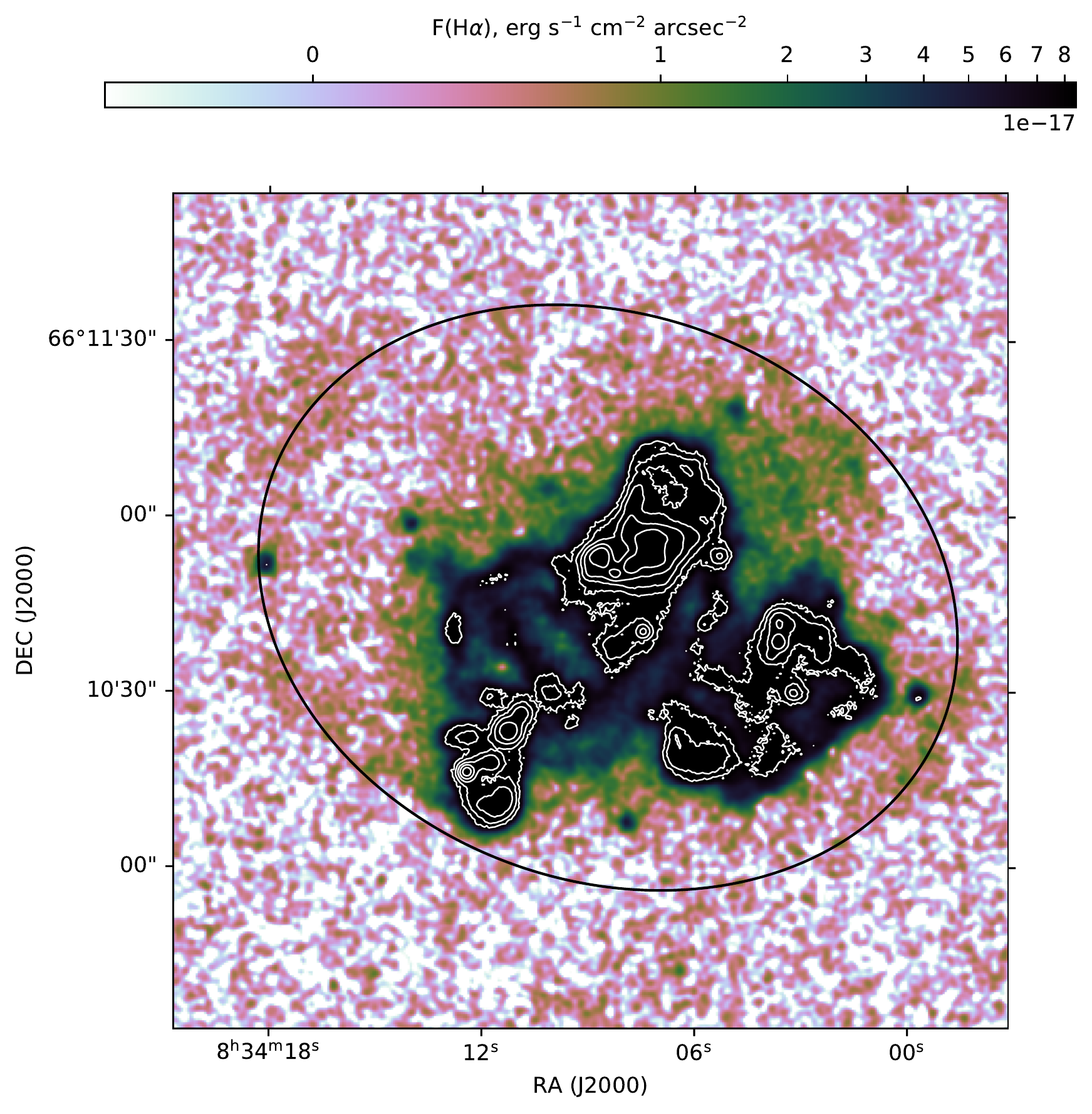}
	\caption{An \Ha image of the galaxy demonstrates the low surface brightness ionised structures outside the regions of ongoing star formation. A black ellipse encompasses the identified \revone{surrounding} shell-like structure. White contours trace the inner \Ha isophotes.}\label{fig:outflow}
\end{figure}

Beside the bright \HII regions and filamentary structures connecting them, we detect very faint 
diffuse emission extending to distances up to 1.2 kpc from the centre of DDO~53 (see Figure~\ref{fig:outflow}). This feature has a surface brightness of about $(2.0-3.8)\times10^{-18}$~erg~s$^{-1}$~cm$^{-2}$~arcsec$^{-2}$ at the southern and north-eastern parts of the galaxy, while they reach a level of about $1.1\times10^{-17}$~erg~s$^{-1}$~cm$^{-2}$~arcsec$^{-2}$ toward the north-west from the galaxy centre. Only the brightest clump of the north-eastern part of this structure is marginally detected in FPI data. Also this clump is well aligned with the extended \HI blue-shifted emission in 21~cm (see Sec.~\ref{sec:HI}). We can encompass all the diffuse ionised gas clumps by an ellipse having position angle of $PA\sim60^\circ$, major semi-axis of $R_{maj}\sim1070$~pc and axis ratio of $R_{min}/R_{maj}\sim0.8$ that corresponds to the inclination of $i\sim35^\circ$, close to the inclination of the galaxy. Hence, the observed diffuse ionised structure \revone{has the form of }
a supershell with size of about 2~kpc. 
We will check further in Sec.~\ref{sec:feedback} if this structure could be a large-scale outflow driven by stellar feedback.

Comparing the \Ha luminosity of the individual star-forming complexes with the deposit of Lyman continuum (LyC) quanta from the identified O stars within them (see Fig.~\ref{fig:regnames} and Sec.~\ref{sec_obs_other}) we may judge the ionisation balance in these regions, as well as for the whole galaxy. From the LEGUS catalogue we can derive the $M_V$ of each star, which can then be translated to the produced number of ionising quanta $Q^0$ according to the model by \cite{Martins2005}. Summing $Q^0$ for all stars within each complex, we find that an available number of $Q^0_*=(5.45, 0.83, 1.76)\times 10^{50}$~s$^{-1}$ for complexes N, SE and SW, respectively. At the same time, the number of quanta required to produce the observed \Ha flux in these complexes could be estimated as $Q^0_\mathrm{H\alpha}\simeq \frac{L\mathrm(H\alpha)}{0.45 h\nu} = (4.07,1.23,1.52)\times 10^{50}$~s$^{-1}$ \citep{Osterbrock2006}. Thus, the identified O stars alone produce enough quanta for gas ionisation within each complex. Note that lower  $Q^0_*/Q^0_\mathrm{H\alpha}$    in the SE complex    is related to the rather arbitrary separation between the complexes SE and N -- some O stars are observed close to the border between them and were counted in the ionisation balance for complex N. Calculating the escape fraction $f_{esc}=1-Q^0_\mathrm{H\alpha}/Q^0_*$ for these complexes together we obtain $f_{esc}=0.16$, similar to that of the SW complex ($f_{esc}=0.14$). These estimates of the escape fraction from the star-forming complexes are lower than those for other dwarf galaxies (e.g., $f_{esc}\sim 0.5-0.7$ for Holmberg~I \citep{Egorov2018} and Holmberg~II \citep{Egorov2017} derived from a similar analysis; $f_{esc}>0.2$ in LMC according to MUSE data in \citealt{McLeod2019}). Such a discrepancy is probably because of the underestimation of the number of O stars, since the LEGUS catalogue we used contains only the stars with the highest-quality photometry (see Sec.~\ref{sec_obs_other}). However even with this number we measure that the total number of available LyC quanta $Q^0_* \sim 8.1\times 10^{50}$~s$^{-1}$ is enough to explain the total \Ha luminosity of the galaxy (resulting to $Q^0_\mathrm{H\alpha} \simeq 7.2\times 10^{50}$~s$^{-1}$), including its diffuse structures like the previously mentioned 2~kpc-sized supershell.

\subsection{Morphology and kinematics of the atomic gas}\label{sec:HI}

\begin{figure}
	\includegraphics[width=\linewidth]{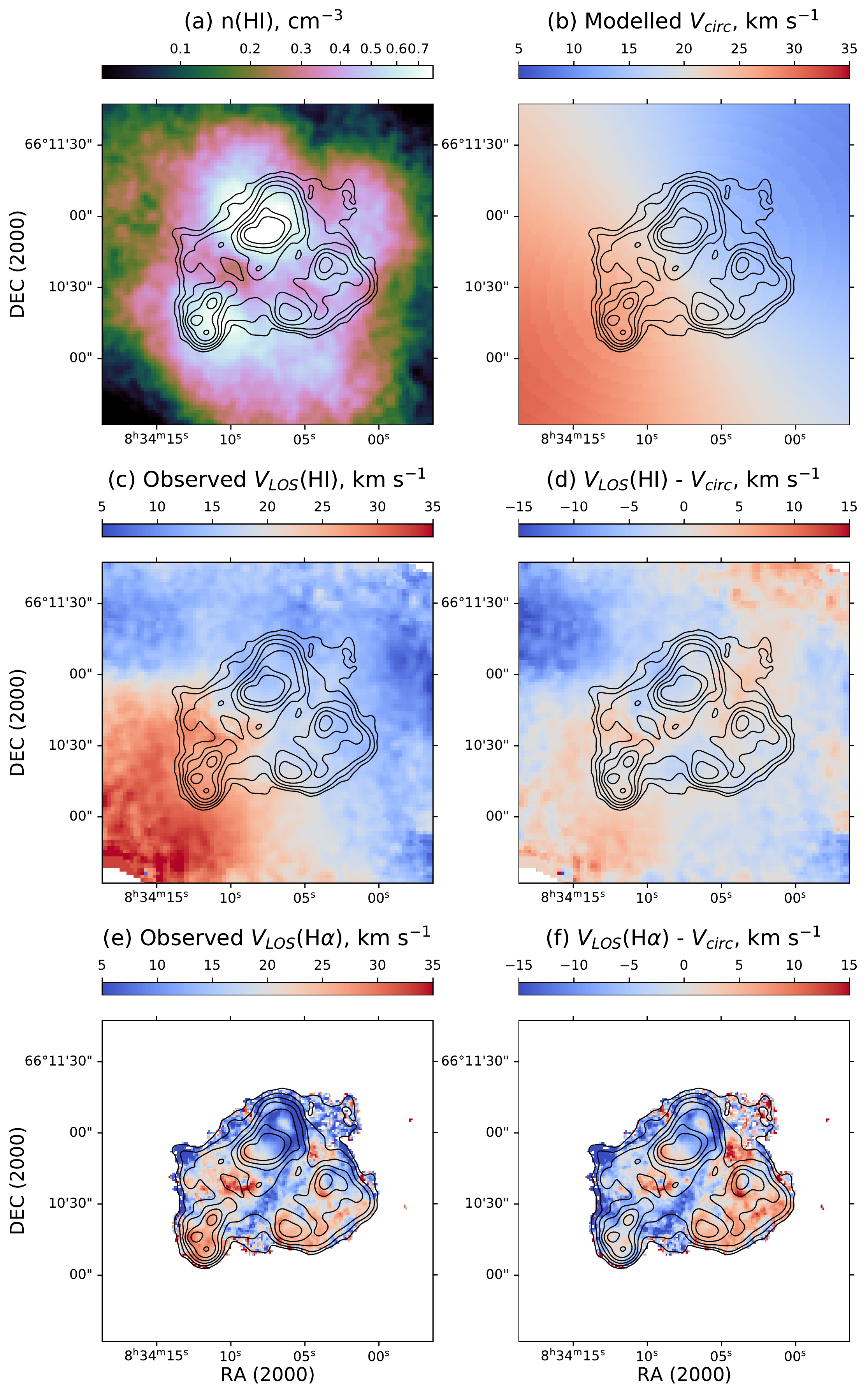}
	\caption{Distribution of the volume density of H~\textsc{i} (a) and the  velocity fields in \HI 21~cm and \Ha lines. Panel (b) shows the modelled velocity field of circular rotation for DDO~53; panels (c) and (d) -- the observed velocity field in \HI and residuals after subtraction of the circular rotation model; panels (e) and (f) -- the same information for the ionised gas in the \Ha line. Contours on each panel are lines of constant \Ha surface brightness.}\label{fig:vel}
\end{figure}

The \HI distribution  in DDO~53 reveals two high-density peaks in the integrated   21 cm map (see Fig.~\ref{fig:vel}a). These peaks coincide with the brightest complexes of star formation -- N and SE (see Fig.~\ref{fig:regnames}). The third star-forming complex -- SW -- resides in the lower density environment, while its brightest nebulae are observed towards the regions with locally enhanced \HI volume density. 

To obtain the distribution of \HI volume density (shown in Fig.~\ref{fig:vel}a) we rely on the natural-weighted \HI data assuming a constant scale height of the \HI disc $h=290$~pc as estimated by \citet{Bagetakos2011}, with the following relation between the column density $N_\mathrm{HI}$ and volume density $n_\mathrm{HI}$:
\begin{equation}
	N_\mathrm{HI}=\int_{-\infty}^{+\infty}n_\mathrm{HI}\exp\left(\frac{-z^2}{2h^2}\right)dz=\sqrt{2\pi}hn_\mathrm{HI}.
	\label{eq:nh}
\end{equation}
The $n_\mathrm{HI}$ is corrected for inclination by multiplying by $\cos(i)$ to take into account the longer path along the line of sight than expected in Eq. (\ref{eq:nh}). The resulting volume density of \HI reaches a value of  $n_\mathrm{HI}\sim0.8$~cm$^{-3}$ in the N and SE complexes, while a typical value of $n_\mathrm{\HI}=0.3-0.5$~cm$^{-3}$ is found in the SW region and outside the bright star-forming complexes. These values translated into a gas mass densities are typical for other nearby dwarfs and the outskirts of massive galaxies \citep[e.g.][]{Abramova2011, Bacchini2019, Bacchini2020}, however the derived values depend on two very uncertain parameters -- disc scale height and inclination. In particular, the inclination angle estimates for DDO~53 vary significantly in the literature -- from $i=27^\circ$ \cite{Oh2011} to $i=56^\circ$ \cite{Boisvert2016}. Here we adopt a value of $i=37^\circ$ measured by \cite{Iorio2017}.

\revone{Using the kinematic centre, position angle $PA_\mathrm{kin}$ and inclination angle (see Table~\ref{tab:ddo53}) derived by \cite{Iorio2017}} 
we reconstruct a model for circular rotation (Fig.~\ref{fig:vel}b) and subtract it from the observed data cubes in both \HI 21 cm and \Ha lines (following the `derotation' procedure described in \citealt{Egorov2014}). The observed line-of-sight \HI velocity field (the first statistical moment of natural-weighted LITTLE THINGS data cube) is shown in Fig.~\ref{fig:vel}c, and the residuals after subtracting the circular motions are shown in Fig.~\ref{fig:vel}d. Similarly, the observed velocity field in \Ha 
and the residuals after subtraction of the circular rotation model are given in Fig.~\ref{fig:vel}e,f (we consider them in Section~\ref{sec:local_kin}).

The recovered simple model of circular rotation well describes  the gas kinematics of the \HI in the central part of DDO~53, while significant residual velocities are observed 
to the north-east from the star-forming regions. \citet{Iorio2017} mentioned the extra emission in the \HI 21 cm line in this area; 
\cite{Hunter2019} found the only one region of strong non-circular motions in DDO~53, which is also located there. Below we try to describe its morphology and kinematics by analysing the \HI data cube after subtraction of the circular motions.

\begin{figure*}
	\includegraphics[width=\linewidth]{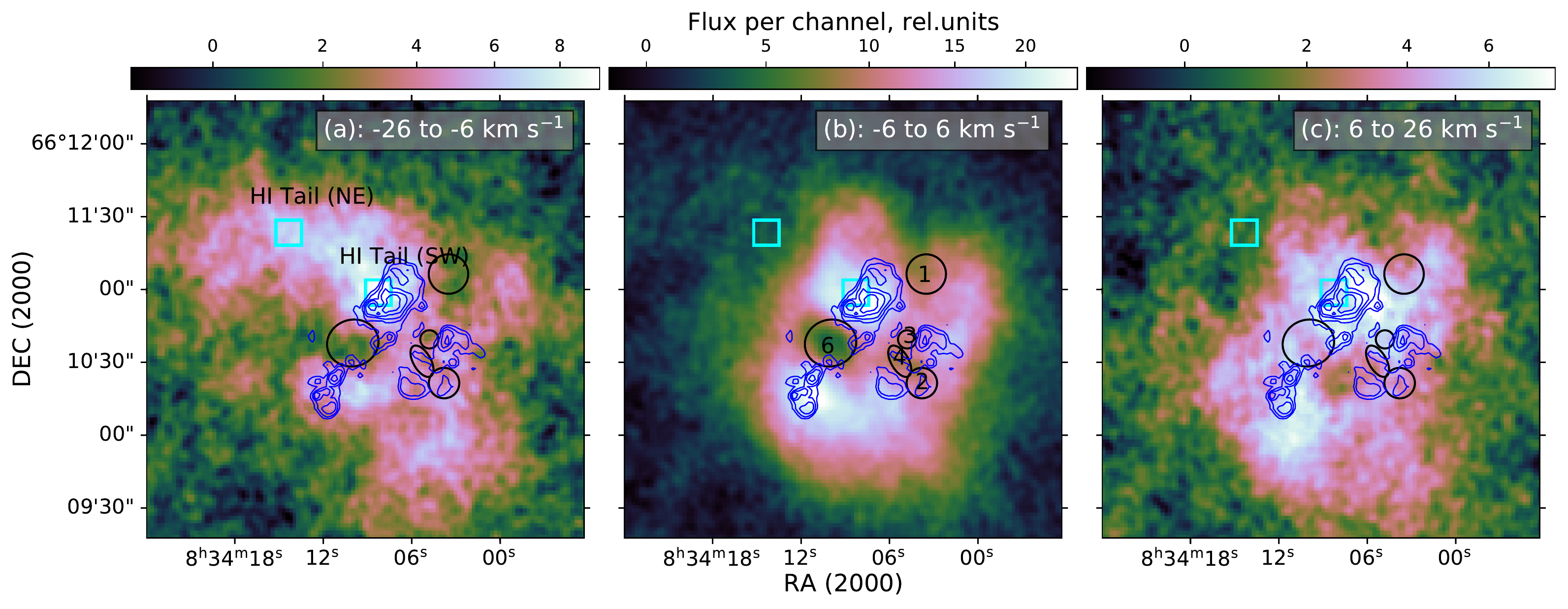}
	
	\includegraphics[width=\linewidth]{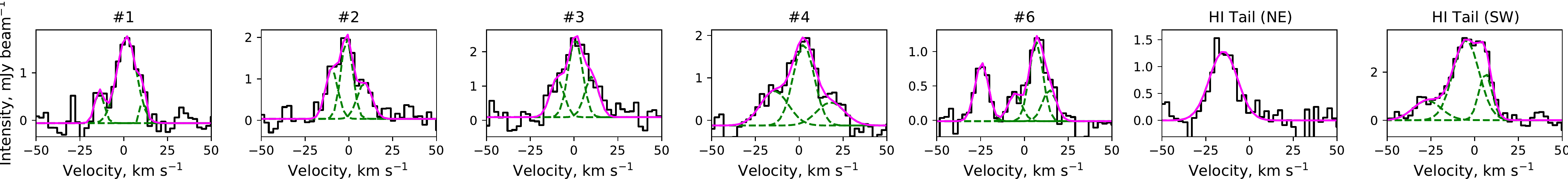}
	\caption{Atomic hydrogen in DDO~53 at different velocities as obtained by integration of several channels in the \HI data cube after extraction of the circular rotation model: (a) blue-shifted gas motions at the velocities of $-26...-6\ \kms$; (b) not shifted motions within $-6...6\ \kms$; (c) red-shifted motions at the velocities of $6...26\ \kms$. Contours on each panel are the lines of constant \Ha surface brightness. Black ellipses show the location of the \HI holes identified by \citet{Pokhrel2020}. The bottom row shows the \HI line profiles extracted towards the centres of the \HI holes and towards the cyan squares (in two places in the \HI tail) and the results of their approximation by 1--3 Gaussians. Individual Gaussian components are shown by dashed green lines, while the  magenta colour is their sum.}\label{fig:hi_cnahhels}
\end{figure*}

We demonstrate in Fig.~\ref{fig:hi_cnahhels} the \HI intensity maps of (a) blue-shifted, (b) not shifted and (c) red-shifted emission. These maps were obtained by integration of the `derotated' \HI data cube within the velocity range of $-26...-6$, $-6...6$ and $6...26\ \mathrm{km\ s^{-1}}$, respectively. 
A long  north-eastern tail in the \HI emission appears only in the blue channels and almost does not contribute to the signal in the central and red-shifted channels. We refer further to this anomalous \HI gas as the `\HI tail'.  

\begin{figure}
	\centering
	\includegraphics[width=\linewidth]{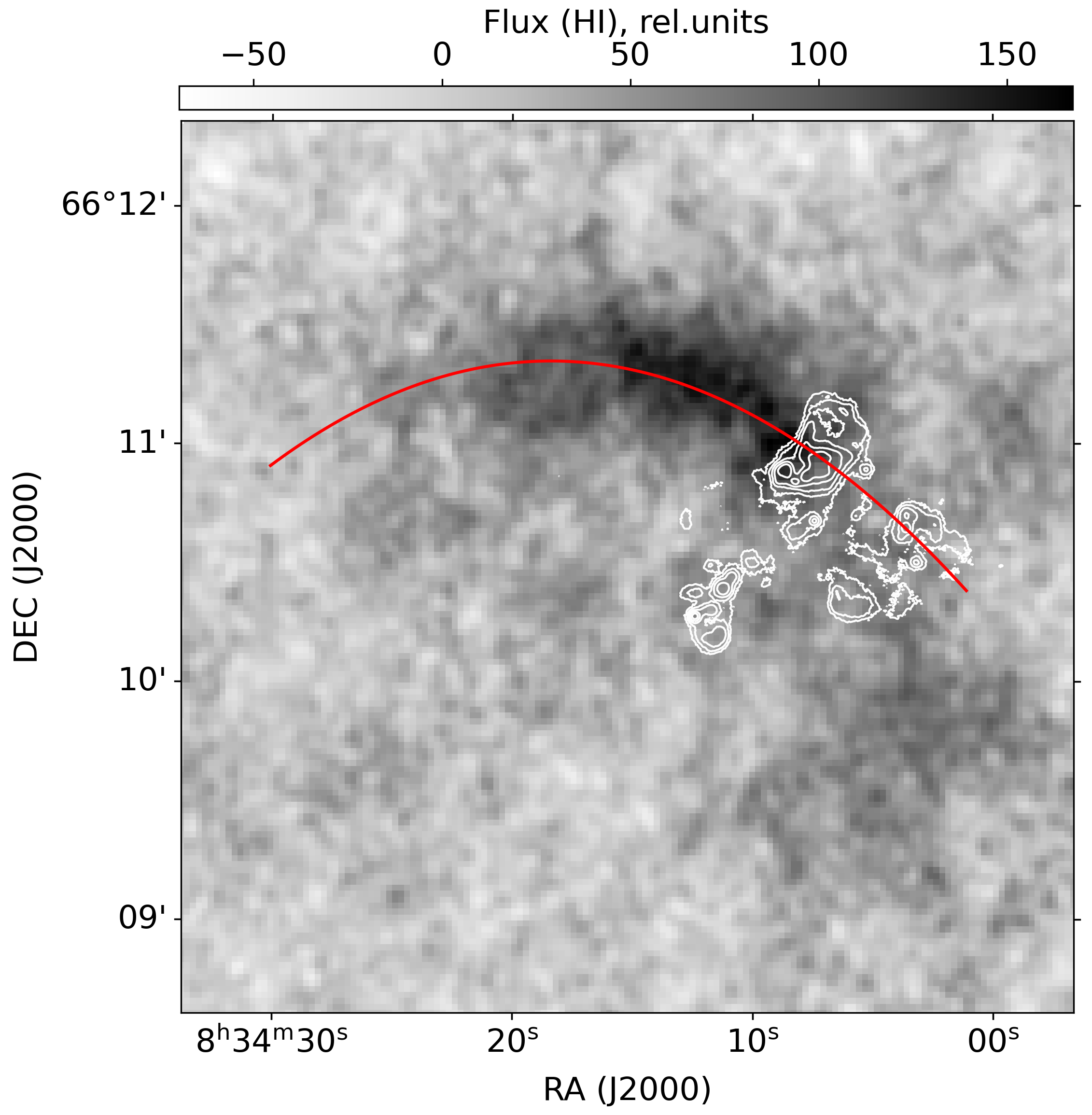}
	\includegraphics[width=\linewidth]{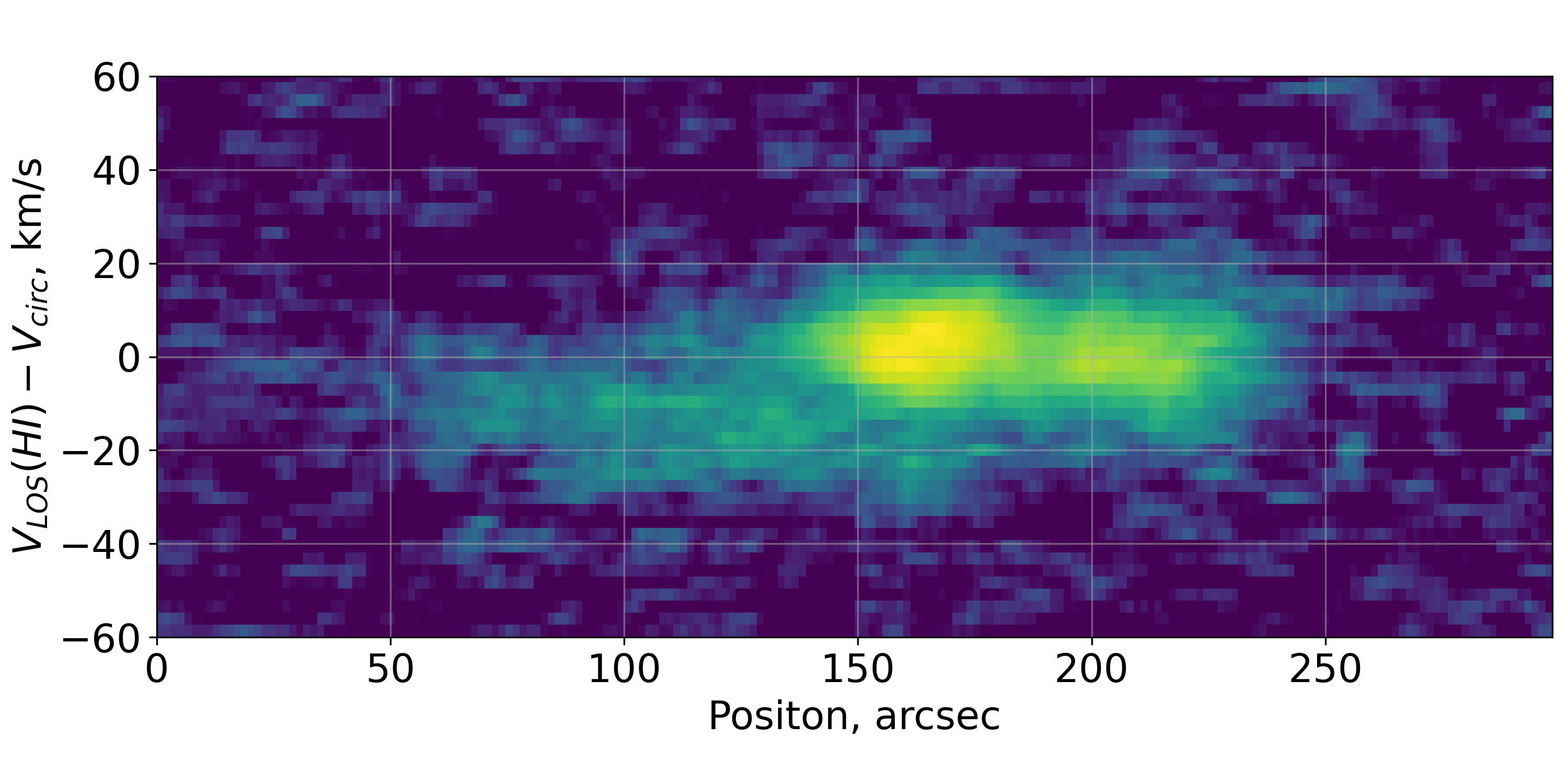}
	\caption{Map of DDO~53 in the blue-shifted \HI channels ($-29...-16\ \mathrm{km\ s^{-1}}$) and the localization of the `position--velocity' (PV) diagram constructed along the \HI tail (top panel). This PV diagram is shown on the bottom panel and demonstrate a clear separation in  velocity space of the \HI tail (left half) from the bulk gas motions (right half). Contours on the top panel are lines of constant \Ha surface brightness.}\label{fig:HI_PV}
\end{figure}

\revone{As it is clearly seen in Fig.~\ref{fig:HI_PV},} the \HI tail is connected with the brightest star-forming region in complex N. The `position -- velocity' diagram shown in the bottom panel is constructed along the whole extent of the tail. It demonstrates that the tail is not connected with the bulk of the galaxy in velocity space. The \HI 21 cm line profile of this tail can be fit by a single Gaussian with a velocity offset of about $20~\kms$, while its contribution in the integrated line profile is still visible at the same velocity toward the region of star formation (see the two right-hand  profiles in Fig.~\ref{fig:hi_cnahhels}). Thus, we may conclude that the gas in the area of non-circular motions of atomic hydrogen mentioned by \cite{Iorio2017} and \cite{Hunter2019} has elongated tail-like form, kinematically detached from the galaxy and probably appears to be an external gas cloud. 
 We discuss this finding further in Section~\ref{sec:accretion}.

The small-scale morphology and kinematics of atomic hydrogen in DDO~53 were previously analysed by \cite{Bagetakos2011} and \cite{Pokhrel2020} who identified several expanding \HI holes (or supershells) despite the rather poor angular resolution of the available \HI 21 cm data (in comparison with the galaxy size). They identified 3 and 7 holes, respectively, having diameters of $170-340$~pc and ages of 6--18~Myr. 
We show the location of the \HI supershells with their names according to \revone{\cite{Pokhrel2020} list (as it is the more complete one)} in Fig.~\ref{fig:hi_cnahhels} with black ellipses. Note that we exclude from consideration supershell~\#5 (it is located much further north from complex N and doesn't correspond to any prominent \HI structure) and \#7 \revone{(which coincides with the \HI tail -- both studies misidentified it as a \HI hole)}.

As expected, all the \HI supershells in DDO~53 are located outside the bright \HII regions, but star-forming regions are observed towards their edges. An exception is the supershell~\#2, which coincides with \HII region~\#3 in the SW complex. Based on the decomposition of the \HI 21 cm line profiles observed towards the central parts of each shell (Fig.~\ref{fig:hi_cnahhels}), we may conclude that clear signs of expansion are indeed visible in all supershells, and the estimated expansion velocities are the same as derived by \cite{Pokhrel2020}. Note that the prominent blue-shifted component at $\sim-24\ \kms$ in the centre of supershell~\#6 in principle could also be related to the gas in the \HI tail given the similarity of the line-of-sight velocities.

\subsection{Small-scale kinematics of the ionised gas}\label{sec:local_kin}

\begin{figure}
	\includegraphics[width=\linewidth]{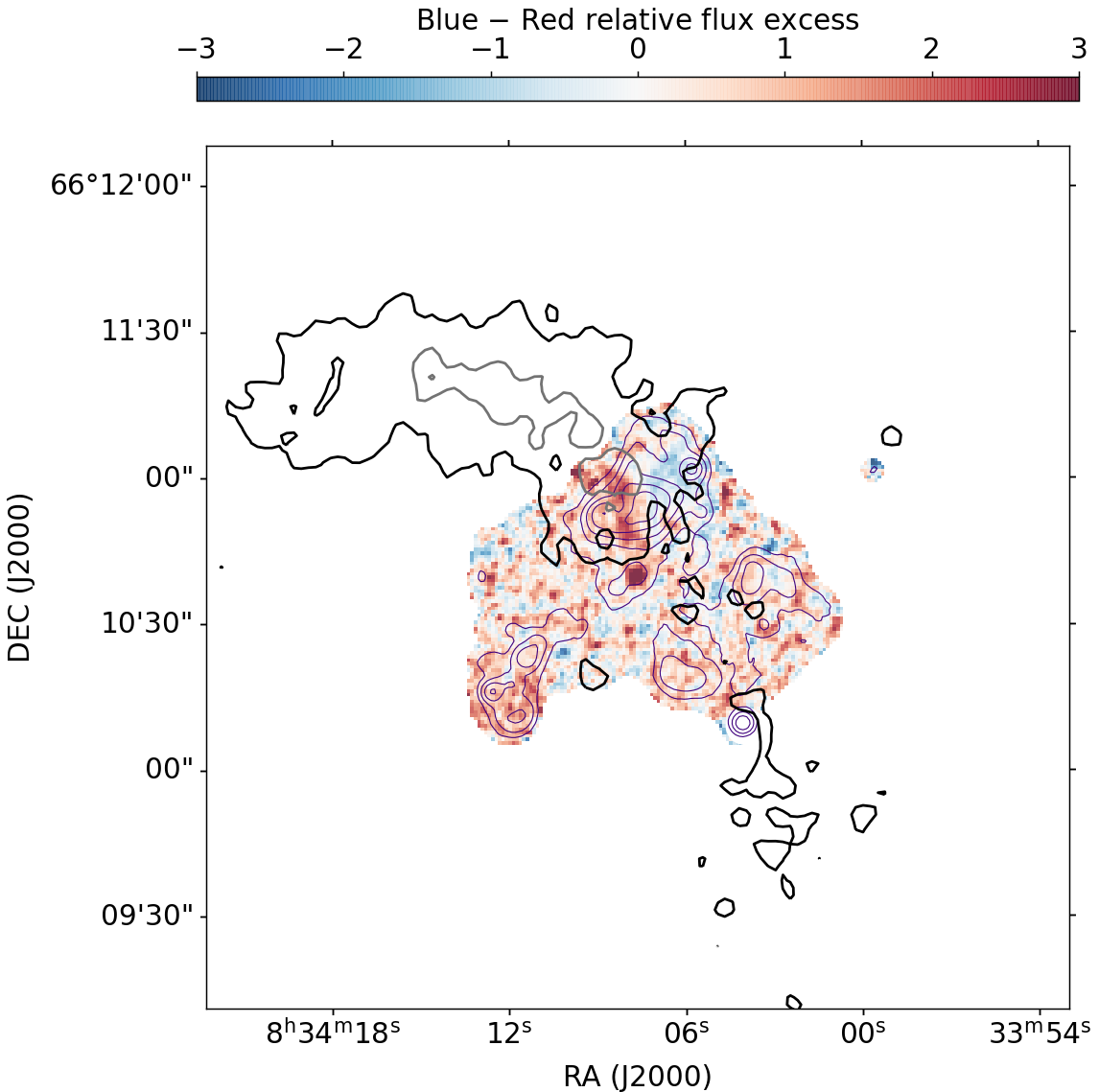}
	\caption{A map of the asymmetry of the \Ha line profile (computed as the difference of the flux excess in the blue- and red-shifted wings after single-component Voigt fitting normalized to the noise level), with overlaid contours of the \HI flux in the blue-shifted channels (black colour). Violet contours correspond to the \Ha flux distribution.}\label{fig:asymm}
\end{figure}

In contrast with the \HI data, high residuals after subtracting the model of circular rotation are still prominent in the velocity field of the ionised gas (panels e and f in Fig.~\ref{fig:vel}). We do not observe any significant correlation of the residuals distribution with the \HI volume density or \Ha surface brightness, however it seems that the red-shifted non-circular motions are more frequent in the SW complex. \revone{Tilted-rings modelling of the ionised gas velocity field using the \Ha FPI data revealed a very similar distribution in the residual velocities \citep{Moiseev2014}. Hence,} we suggest that the significant non-circular motions result from the stellar feedback influencing scales smaller than the resolution of \HI data rather than from the general differences between atomic and ionised gas kinematics. Since the effects of the galaxy rotation contribute to the observed gas kinematics of DDO~53 to the same extent as the stellar feedback, we further use the `derotated' data for the \Ha kinematics to concentrate mostly on the local gas motions.

The observed shape of the \Ha line in the FPI data is highly symmetric and well described by a single Voigt profile in most of the galaxy, however noticeable asymmetry is present in  complex N where the \HI tail connects with the brightest \HII regions. To explore this, we fit the whole data cube with a single Voigt profile and analyse the residuals. As a proxy for the line asymmetry we use the \revone{difference of the flux excess in the blue- and red-shifted wings} 
normalized to the noise level at the corresponding pixel. The resulting map is shown in Fig.~\ref{fig:asymm}. The reddest colour shows the regions where the intensity of the red-shifted component exceeds the 3$\sigma$ level, while the bluest colour corresponds to the same for the blue-shifted component relative to the result of a single-component fitting. The only extended region with a large contribution of the shifted component to the integrated \Ha line profile is elongated from the \HII region \#24 (see Fig.~\ref{fig:regnames}) towards the brightest regions \#31, 33, 35 and coincides well with the direction of the \HI tail (shown by black contours). Hence, we may suggest that the shifted component in the \Ha line and the \HI tail are related to each other.

\begin{figure}
	\includegraphics[width=\linewidth]{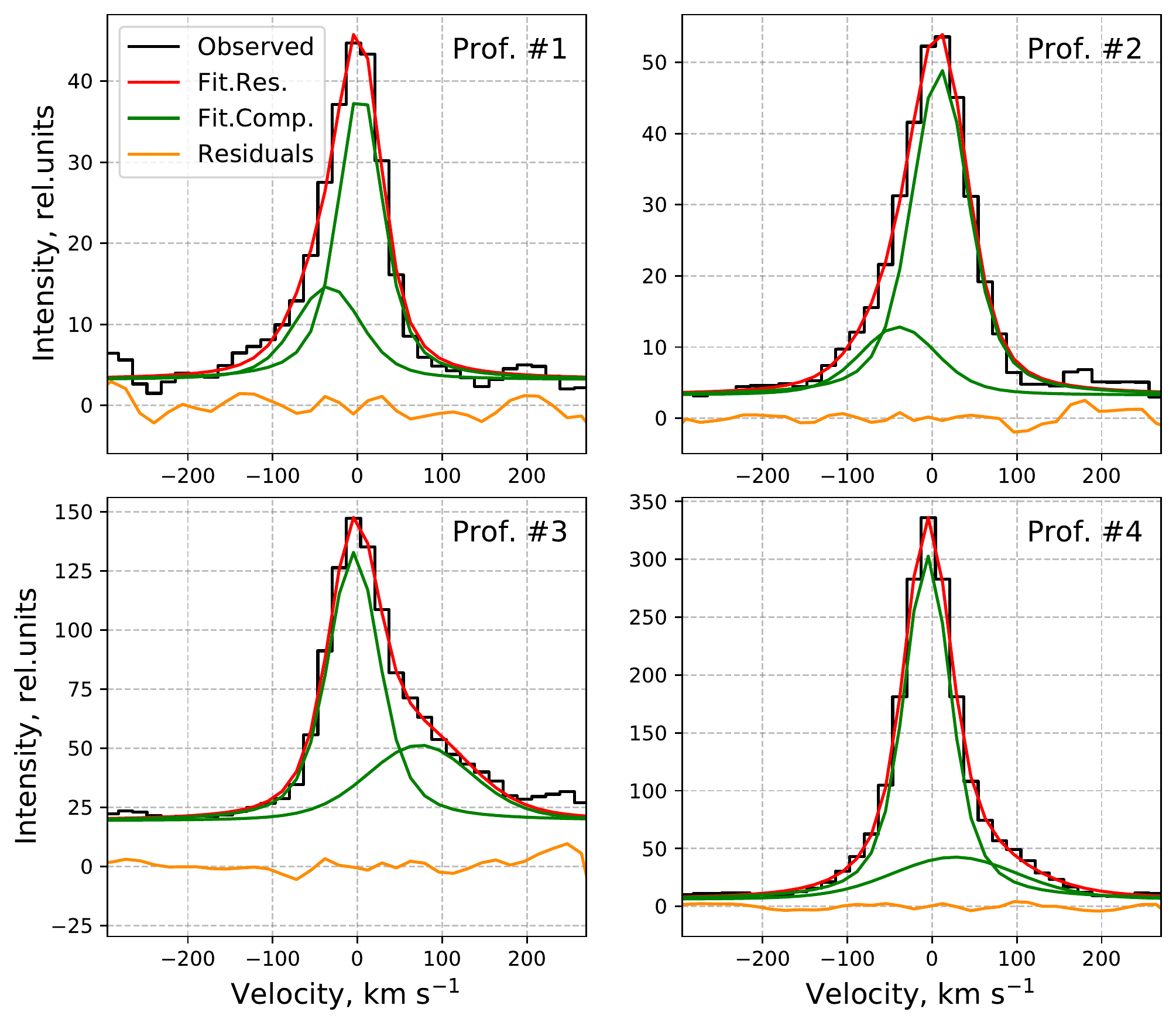}
	\caption{Example of \Ha line profiles extracted from FPI data within the regions shown by green squares in Fig.~\ref{fig:isigma}c, and the results of their fitting with 1- or 2-component Voigt functions. Profile ~\#1 corresponds to the \HII region~\#2 and appears as single component line profile with $\sigma$(H$\alpha$)$=18.5 \kms$ that is typical for DDO~53. Other profiles are from the only three regions in the galaxy where kinematically distinct components were detected. The most prominent of these (profile~\#2) corresponds to region~\#24 discussed in Sec.~\ref{sec:spectra_res}. }\label{fig:FPI_profiles}
\end{figure}

\begin{figure*}
	\includegraphics[width=0.95\linewidth]{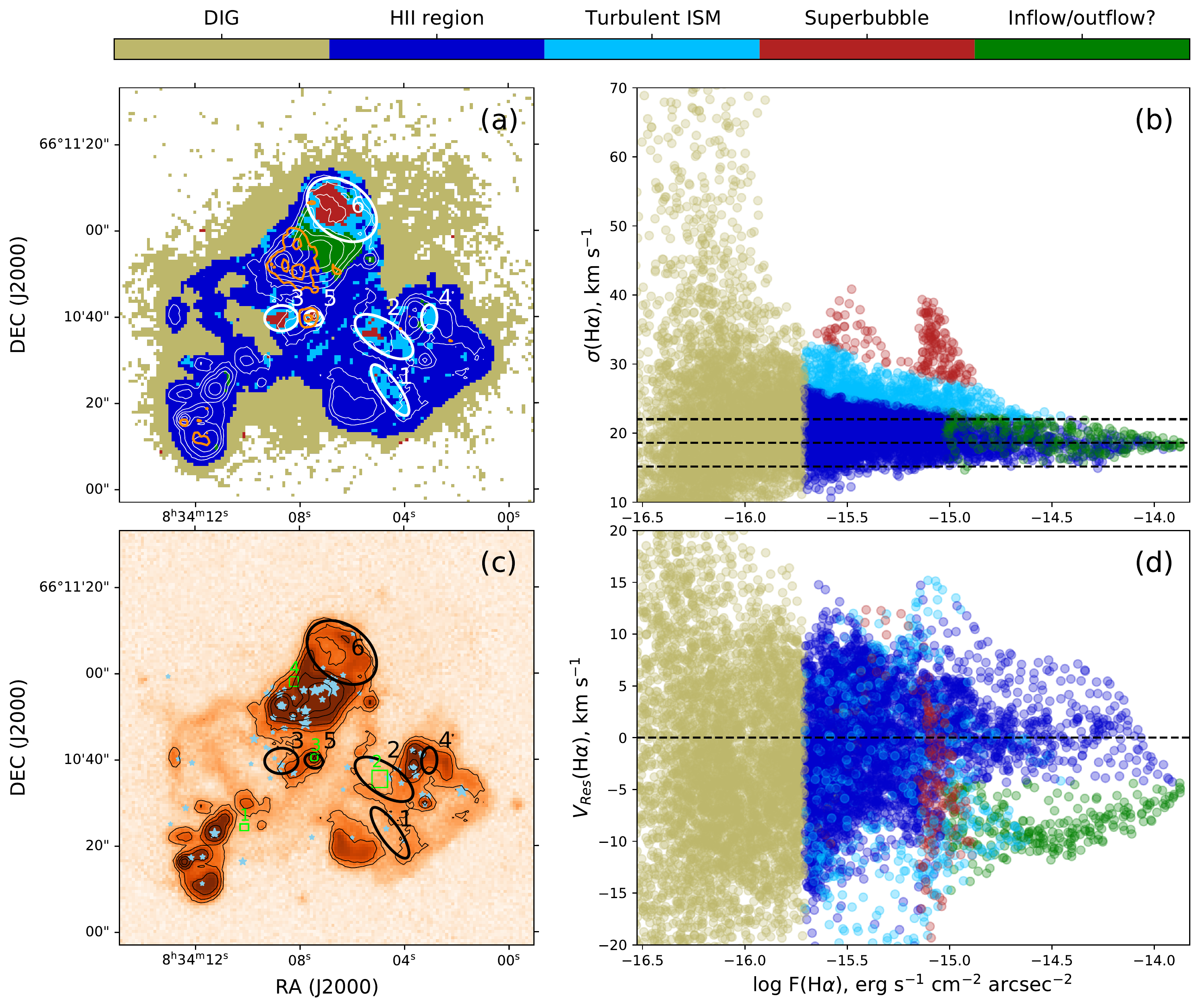}
	\caption{Analysis of the \Ha velocity dispersion $\sigma$(H$\alpha$) in the FPI data. \textbf{Panel (a)} -- classification map according to the diagrams presented on panels (b) and (c). Different colours encode several types of regions denoted on the colour-bar (colours are the same in panels (a), (b), (d). White ellipses are the superbubble candidates that have elevated $\sigma$(H$\alpha$) and are selected with \textsc{astrodendro}. Orange contours correspond to the \Ha line profile asymmetry distribution as shown in Fig.~\ref{fig:asymm}. White contours show the distribution of \Ha surface brightness.  \textbf{Panel (b)} --  `I--$\sigma$' diagram of the \Ha line showing the pixel-by-pixel dependence of the $\sigma$(H$\alpha$) on the logarithm of the line flux F(H$\alpha$). Olive colours correspond to the DIG and is selected as having F(H$\alpha$) less than the median value for the galaxy. Blue colours trace the areas with $\sigma$(H$\alpha$) within the normally distributed values for the corresponding flux around the flux-weighted mean value $\sigma_m$(H$\alpha)=18.6\pm3.5 \kms$ (shown by the dashed lines). Cyan colours are the regions having $\sigma$(H$\alpha$) beyond this normal distribution, and red colours are peaks in the diagram having $\sigma$(H$\alpha$) larger than found for a normal distribution with twice the standard deviation around $\sigma_m$(H$\alpha$).  \textbf{Panel (c)} --  \Ha image of the galaxy with overlaid ellipses from panel (a) and location of the main sequence O-stars. Black contours trace the \Ha surface brightness.   \textbf{Panel (d)} -- pixel-by-pixel analysis of the line-of-sight residual velocity V$_{res}$(H$\alpha$) versus $\log$ F(H$\alpha$). The pixels with blue-shifted V(H$\alpha$) and  $\sigma$(H$\alpha$)  normally distributed around $\sigma_m$(H$\alpha$) are denoted by green colours on panels (a-b), (d). Note that \revone{on panel (d)} the points from such blue-shifted regions smoothly turn into the points related to the superbubble SB-6.}\label{fig:isigma} 
\end{figure*}

Visual inspection of the FPI data cube reveal only four regions where the \Ha line profile could be reliably decomposed onto two kinematically distinct components, and all these regions are present on the map of \Ha asymmetry as red or blue spots that confirm the result presented in Fig.~\ref{fig:asymm}. Examples of the line profiles towards such regions and the results of their decomposition are given in Fig.~\ref{fig:FPI_profiles}. The exact borders over which the profiles were integrated are shown in Fig.~\ref{fig:isigma}c with a green colour. In all these cases the second component is broader than the central one. Several other spots in the map of \Ha line asymmetry also could be in principle decomposed into two components (in particular -- in the shell to the north-west of the brightest \HII region), but the S/N or the separation between components is too low there to get a reliable result.

\begin{table*}
	\caption{Derived parameters of the superbubbles selected from the I-$\sigma$ diagram in Fig.~\ref{fig:isigma}}\label{tab:superbubbles}
	\begin{scriptsize}
		\begin{tabular}{cccccccccc}
			\hline
			\# & RA (J2000)& Dec (J2000) & $D_{min}$, pc & $D_{maj}$, pc & PA &  $\sigma_{int},\ \kms$ & $V_{exp},\ \kms$ & Age, Myr & $\dot{E}_{kin}$, $10^{36}$~erg~s$^{-1}$ \\
			\hline
			SB-1 & 128.51897 & 66.17306 & 80 & 251 & 35 & 21 & 18 & 2.4 & 1.5 \\
			SB-2 & 128.51990 & 66.17650 & 127 & 275 & 57 & 23 & 21 & 2.6 & 4.6 \\
			SB-3 & 128.53627 & 66.17770 & 106 & 138 & 92 & 28 & 33 & 1.1 & 6.8 \\
			SB-4 & 128.51268 & 66.17772 & 64 & 108 & 178 & 23 & 24 & 1.1 & 1.2 \\
			SB-5 & 128.53108 & 66.17770 & 60 & 77 & 72 & 26 & 30 & 0.7 & 1.6 \\
			SB-6 & 128.52660 & 66.18468 & 224 & 321 & 52 & 25 & 28 & 2.9 & 20.5 \\
			\hline
		\end{tabular}
	\end{scriptsize}
\end{table*}

From the analysis of the \Ha data cube, \cite{Moiseev2012} previously identified three regions of high velocity dispersion of ionised gas (see their fig.~1) which were considered to be probable unresolved expanding superbubbles, however the origin of such supersonic motions remained unknown. Here we reproduce their analysis using a slightly different criteria and selection algorithm. For that we use the so-called $I-\sigma$ diagrams, showing the dependence of the velocity dispersion $\sigma$(H$\alpha$) on the logarithm of the \Ha line flux in each corresponding pixel (see Fig.~\ref{fig:isigma}b). Such diagrams were introduced by \cite{MunozTunon1996} and further improved by \cite{Moiseev2012}. Different components of the ionised ISM occupy different places on the $I-\sigma$ diagram. First, we exclude all the pixels with \Ha flux below the median value for DDO~53 as probably corresponding to the DIG --- $\Sigma(\mathrm{H\alpha}) \simeq 10^{39}$ erg s$^{−1}$ kpc$^{−2}$ that was considered by \citet{Zhang2017} as demarcating surface brightness between DIG and \HII regions). \HII regions are a dominant features on the rest of the diagram and appear as having relatively high \Ha flux and $\sigma$(H$\alpha$) close to the flux-weighted mean value in the galaxy ($\sigma_m=18.6\pm3.4\ \kms$). We classify a pixel as related to an \HII region (blue colour) if its $\sigma$(H$\alpha$) lies within the normal distribution around $\sigma_m$ \revone{at each $I$ value}. As follows from the classification map (panel a) and its comparison with \Ha map (panel c), the prominent faint filaments that physically are not \HII regions, but do not demonstrate elevated velocity dispersion, are also classified as \HII regions following our criteria. Finally, the regions having  $\sigma$(H$\alpha$) higher than defined by a normal distribution around $\sigma_m$ with standard deviation equal to 1 or 2 times the uncertainties of $\sigma_m$ (3~$\kms$) are considered either turbulent ISM (cyan colour) or superbubble candidates (red colour), respectively. \revone{According to \cite{MunozTunon1996}, the expanding superbubbles should appear on the $I-\sigma$ diagram as  triangle-shaped features, as they appear in  Fig.~\ref{fig:isigma}b.}

\revone{In Fig.~\ref{fig:isigma}d we show the distribution of the FPI pixels on the plane of 
the residual line-of-sight velocity $V_{Res}$(H$\alpha$) (after circular rotation subtraction) 
versus $\log F$(H$\alpha$). The colours are the same as in the $I-\sigma$ diagram, but here we also introduce the area corresponding to the well detached blue-shifted ionised gas motions that do not show elevated velocity dispersion (encoded in green colours on both of these diagrams and on panel a). As follows from Fig.~\ref{fig:isigma}a, this area is observed towards the brightest \HII regions in complex N, exactly in the place where the \HI tail is observed. It is connected with the area of large \Ha line profile asymmetry. 
 The absolute offset ($\sim 10\ \kms$) of these motions is slightly lower than for the \HI tail, but they still might be related with each other (see Sec.~\ref{sec:accretion})}

 We apply the \textsc{astrodendro} routine to select all spatially-connected pixels classified as turbulent ISM or superbubbles and having size not less than angular resolution of the data. 
As a result, we identify 6 regions approximated by ellipses; their numbers are given on panels a and c. Among them, four regions contain the extended areas classified as \revone{expanding} superbubbles (SB-2, 3, 5, 6), while the regions SB-1 and SB-4 have lower maximal velocity dispersion. We consider all the selected regions as superbubble candidates. Note that while only region SB-6 demonstrates a clear shell-like morphology, other regions (except the most compact SB-5) are still encircled by the \Ha filamentary structures or \HII regions that are typical for superbubbles in an inhomogeneous medium. 

Three of the \revone{expanding} superbubbles coincide with the regions selected by \cite{Moiseev2012}, while  SB-5 was not identified previously. The latter region appears as a small \HII region \#24 (according to Fig.~\ref{fig:regnames}), demonstrating the most prominent asymmetry of the \Ha line profile (Fig.~\ref{fig:asymm} and also orange contours in Fig.~\ref{fig:isigma}a) that could be reliably decomposed onto two kinematically distinct components, one of which is broad and red-shifted (prof.~\#3 in Fig.~\ref{fig:FPI_profiles}). 
Among the rest of the selected superbubble candidates, only SB-2 demonstrates a multi-component \Ha line profile at the current resolution (prof.~\#2). As follows from Fig.~\ref{fig:isigma}c, all of the selected regions do not contain numerous OB stars, yet a few of them are encircled by these superbubble candidates. Such a picture was observed earlier in other dIrr galaxies \cite[e.g.][]{Egorov2017}. In some cases (SB-4 and SB-6) the elevated velocity dispersion are observed just beyond the prominent star clusters located within the bright \HII regions. We suggest that such regions appear to be superbubbles expanding into the lower density environment outside the star-forming regions under the influence of the energy and momentum ejection of the OB stars (and supernovae) residing in the bright \HII regions. 

Because the observed \Ha line profile doesn't show a clear separation into the components from the approaching and receding sides, we cannot precisely measure the expansion velocities $V_{exp}$ of the identified superbubbles. However, we can still estimate it from the velocity dispersion $\sigma_{int}$ derived from the integrated spectrum of the whole superbubble region.  Smirnov-Pinchukov \& Egorov (2021, \revone{in press})   demonstrated that $V_{exp}$ can be expressed as  $V_{exp} = k(\sigma_{int}^2 - \sigma_{m}^2)^\alpha + V_0$ with the coefficients $k$, $\alpha$ and $V_0$ dependent on the mean value of velocity dispersion outside a superbubble $\sigma_{m}$ and on the given FWHM of the instrumental profile\footnote{In case of DDO~53 we utilised FPI $k=3.14$, $\alpha=0.4$, $V_0=-3$}.
We list the derived values for each superbubble in Table~\ref{tab:superbubbles} together with their centres (RA, Dec), minor and major diameters $D_{min}$, $D_{maj}$ (measured as the size of the ellipses in Fig.~\ref{fig:isigma}, which enclose 80 per cent of the area with elevated velocity dispersion for each region), position angle PA, kinematic age and kinetic energy consumption $\dot E_{kin}$.  The kinematic age is derived assuming the \cite{Weaver1977} model of evolution of superbubbles from which we could derive $t=0.6 R/V_{exp}$. The effective radius is derived as $R=0.5 \sqrt{D_{min}\times D_{maj}}$. The kinetic energy retained in the superbubbles is estimated as $\dot E_{kin} = 0.5 \dot M_{sh} V_{exp}^2 = 2\pi n \mu m_\mathrm{H} R^2 V_{exp}^3 $ assuming $\mu=1.4$ (taking into account the helium contribution) and volume density $n=0.4$~cm$^{-3}$ (see Fig.~\ref{fig:vel}). As follows from  Table~\ref{tab:superbubbles}, the identified superbubbles are relatively young and have ages of 1--3 Myr. That is consistent with the estimate made by \cite{Moiseev2012} who roughly assumed the maximal velocity dispersion as a proxy of the expansion velocity. These values are also consistent with the estimates of the age of the \HII regions derived in Section~\ref{sec:spectra_res}. Comparing the measured values of $\dot E_{kin}$ with the mechanical wind luminosity of single O star ($L_w \sim 0.8\times10^{36}$~erg~s$^{-1}$ for an O3V star at a metallicity of $Z=0.2Z_\odot$, according to \citealt{Smith2002}), we may conclude that even the few OB stars residing in the centre or the edges of SB-1, SB-2, SB-3, SB-4 produce enough energy to drive the expansion of the superbubbles, especially assuming that such a calculation doesn't take into account contribution of the radiation pressure and of the possible supernovae explosions. An additional source of energy beyond the winds of the OB stars (perhaps, supernovae, but see further discussion on the external gas flow and shocks in Sec.~\ref{sec:accretion}) are required to drive the expansion of SB-6.

%
%
%
%
%

\subsection{Gas excitation and metallicity of the \HII regions and superbubbles}\label{sec:spectra_res}

From our long-slit and IFU spectroscopic observations we were able to derive the excitation state and physical properties of more than half of 37 identified \HII regions in DDO~53 and also in some diffuse emission structures related to the identified superbubbles. We analysed the integrated spectra of those parts of the \HII regions crossed by the slit, or the integrated spectra over the whole extent of the \HII region (within the borders shown in Fig.~\ref{fig:regnames}) in the case of the IFU observations. The results of the analysis are given in Table~\ref{tab:ls_fluxes}. All the listed flux ratios were corrected for reddening estimated from the measured Balmer decrement as described in Sec.~\ref{sec_obs_ls}; the corresponding colour excess $E(B-V)$ for each region are given in Table~\ref{tab:ls_fluxes}. 
We provide also measurements of the equivalent width of the H$\beta$ line, EW(H$\beta$), the age of the \HII regions as estimated from the EW(H$\beta$) according to the model of \cite{Levesque2013}, and the oxygen abundance $\mathrm{12+\log(O/H)}$ measured using two methods (see below).

\begin{table*}
	\caption{Results of the analysis of the spectra for individual \HII regions according to  Fig.~\ref{fig:regnames}. For each region we provide the source of the data (IFU or long slit with corresponding position angle) and the ranges within with the spectra were integrated (for IFU -- the elliptical aperture shown in Fig.~\ref{fig:regnames}). Regions with names that start with SB are the superbubbles identified in Section~\ref{sec:local_kin} (see Fig.~\ref{fig:isigma}).}\label{tab:ls_fluxes}
	\begin{scriptsize}
		
		\begin{tabular}{lccccccc}
			\hline
			Region & $\mathrm{1}$ & $\mathrm{2}$ & $\mathrm{2}$ & $\mathrm{3}$ & $\mathrm{5}$ & $\mathrm{7}$ & $\mathrm{7}$ \\ 
			\hline
			Slit & IFU & PA323 & IFU & PA211 & PA198 & PA323 & IFU \\ 
			Pos.(arcsec) & -- & -32 -- -27 & -- & 25 -- 31 & 16 -- 27 & -21 -- -16 & -- \\ 
			E(B-V) & $0.00 \pm 0.04$ & $0.37 \pm 0.03$ & $0.00 \pm 0.03$ & $0.21 \pm 0.12$ & $0.03 \pm 0.02$ & $0.18 \pm 0.04$ & $0.09 \pm 0.04$ \\ 
			{[O \textsc{ii}] 3727,3729}/H$\beta$ & $1.42 \pm 0.16$ & $-$ & $0.21 \pm 0.23$ & $-$ & $2.27 \pm 0.67$ & $-$ & $2.36 \pm 0.37$ \\ 
			{[O \textsc{iii}] 5007}/H$\beta$ & $2.41 \pm 0.11$ & $4.45 \pm 0.11$ & $3.74 \pm 0.14$ & $0.56 \pm 0.11$ & $1.51 \pm 0.03$ & $1.56 \pm 0.08$ & $2.61 \pm 0.14$ \\ 
			{[N \textsc{ii}] 6583}/H$\alpha$ & $0.027 \pm 0.010$ & $0.011 \pm 0.003$ & $< 0.003$ & $< 0.055$ & $0.034 \pm 0.003$ & $0.035 \pm 0.006$ & $0.024 \pm 0.009$ \\ 
			{[S \textsc{ii}] 6717,6731}/H$\alpha$ & $0.090 \pm 0.009$ & $0.030 \pm 0.003$ & $0.072 \pm 0.014$ & $0.193 \pm 0.029$ & $0.120 \pm 0.003$ & $0.131 \pm 0.009$ & $0.106 \pm 0.011$ \\ 
			{[Ar \textsc{iii}] 7136}/H$\alpha$ & $-$ & $-$ & $-$ & $-$ & $0.016 \pm 0.002$ & $0.033 \pm 0.017$ & $-$ \\ 
			EW(H$\beta$) & $114.9 \pm 9.9$ & $72.5 \pm 6.3$ & $199.3 \pm 20.8$ & $19.4 \pm 7.4$ & $61.6 \pm 3.4$ & $60.6 \pm 9.2$ & $56.4 \pm 3.4$ \\ 
			Age (Myr) & $4.3 \pm 0.1$ & $4.8 \pm 0.1$ & $3.4 \pm 0.3$ & $6.5 \pm 0.7$ & $4.9 \pm 0.1$ & $5.0 \pm 0.1$ & $5.0 \pm 0.1$ \\ 
			$\mathrm{12+\log(O/H)_S}$ & $7.71 \pm 0.10$ & $7.76 \pm 0.08$ & $-$ & $-$ & $7.61 \pm 0.03$ & $7.63 \pm 0.06$ & $7.71 \pm 0.11$ \\ 
			\hline
			\hline
			Region & $\mathrm{8}$ & $\mathrm{10}$ & $\mathrm{11}$ & $\mathrm{14}$ & $\mathrm{14}$ & $\mathrm{15}$ & $\mathrm{18}$ \\ 
			\hline
			Slit & PA323 & PA259 & PA323 & PA211 & PA259 & PA259 & PA259 \\ 
			Pos.(arcsec) & -24 -- -22 & 46 -- 56 & -9 -- -5 & 17 -- 21 & 38 -- 43 & 20 -- 29 & 6 -- 12 \\ 
			E(B-V) & $0.14 \pm 0.07$ & $0.07 \pm 0.05$ & $0.00 \pm 0.13$ & $0.00 \pm 0.10$ & $0.09 \pm 0.02$ & $0.08 \pm 0.06$ & $0.04 \pm 0.03$ \\ 
			{[O \textsc{iii}] 5007}/H$\beta$ & $0.86 \pm 0.10$ & $0.80 \pm 0.06$ & $0.12 \pm 0.08$ & $0.39 \pm 0.09$ & $1.02 \pm 0.02$ & $0.29 \pm 0.09$ & $0.81 \pm 0.04$ \\ 
			{[N \textsc{ii}] 6583}/H$\alpha$ & $0.051 \pm 0.012$ & $0.056 \pm 0.014$ & $0.072 \pm 0.021$ & $< 0.044$ & $0.031 \pm 0.004$ & $0.052 \pm 0.008$ & $0.039 \pm 0.006$ \\ 
			{[S \textsc{ii}] 6717,6731}/H$\alpha$ & $0.145 \pm 0.018$ & $0.117 \pm 0.014$ & $0.226 \pm 0.030$ & $0.237 \pm 0.030$ & $0.128 \pm 0.004$ & $0.238 \pm 0.012$ & $0.198 \pm 0.009$ \\ 
			{[Ar \textsc{iii}] 7136}/H$\alpha$ & $-$ & $-$ & $-$ & $-$ & $0.008 \pm 0.004$ & $-$ & $-$ \\ 
			EW(H$\beta$) & $24.9 \pm 4.6$ & $54.4 \pm 13.0$ & $15.6 \pm 4.5$ & $38.2 \pm 8.3$ & $112.4 \pm 9.4$ & $35.8 \pm 4.9$ & $34.0 \pm 3.1$ \\ 
			Age (Myr) & $6.2 \pm 0.3$ & $5.0 \pm 0.3$ & $6.9 \pm 0.4$ & $5.4 \pm 0.3$ & $4.3 \pm 0.1$ & $5.5 \pm 0.2$ & $5.6 \pm 0.2$ \\ 
			$\mathrm{12+\log(O/H)_S}$ & $7.53 \pm 0.10$ & $7.52 \pm 0.08$ & $-$ & $-$ & $7.43 \pm 0.04$ & $7.18 \pm 0.09$ & $7.44 \pm 0.05$ \\ 
			\hline
			\hline
			Region & $\mathrm{20}$ & $\mathrm{22}$ & $\mathrm{23}$ & $\mathrm{24}$ & $\mathrm{27}$ & $\mathrm{31}$ & $\mathrm{31}$ \\ 
			\hline
			Slit & PA259 & PA211 & PA259 & PA198 & PA211 & PA198 & PA323 \\ 
			Pos.(arcsec) & -14 -- -5 & 9 -- 16 & -22 -- -17 & -4 -- 2 & -4 -- 3 & -24 -- -19 & 12 -- 16 \\ 
			E(B-V) & $0.13 \pm 0.07$ & $0.09 \pm 0.06$ & $0.08 \pm 0.07$ & $0.30 \pm 0.02$ & $0.05 \pm 0.08$ & $0.01 \pm 0.02$ & $0.07 \pm 0.01$ \\ 
			{[O \textsc{ii}] 3727,3729}/H$\beta$ & $-$ & $-$ & $-$ & $-$ & $-$ & $1.20 \pm 1.26$ & $1.59 \pm 0.55$ \\ 
			{[O \textsc{iii}] 5007}/H$\beta$ & $0.39 \pm 0.08$ & $0.49 \pm 0.07$ & $0.73 \pm 0.07$ & $0.94 \pm 0.02$ & $0.23 \pm 0.07$ & $1.40 \pm 0.04$ & $1.58 \pm 0.02$ \\ 
			{[N \textsc{ii}] 6583}/H$\alpha$ & $0.052 \pm 0.016$ & $0.044 \pm 0.010$ & $0.031 \pm 0.011$ & $0.026 \pm 0.003$ & $< 0.043$ & $0.041 \pm 0.005$ & $0.029 \pm 0.002$ \\ 
			{[S \textsc{ii}] 6717,6731}/H$\alpha$ & $0.220 \pm 0.022$ & $0.238 \pm 0.017$ & $0.152 \pm 0.013$ & $0.068 \pm 0.003$ & $0.241 \pm 0.025$ & $0.165 \pm 0.007$ & $0.123 \pm 0.003$ \\ 
			{[Ar \textsc{iii}] 7136}/H$\alpha$ & $-$ & $-$ & $-$ & $0.006 \pm 0.002$ & $-$ & $0.012 \pm 0.005$ & $0.012 \pm 0.001$ \\ 
			EW(H$\beta$) & $24.7 \pm 7.5$ & $20.9 \pm 4.0$ & $58.2 \pm 10.7$ & $39.7 \pm 2.1$ & $32.4 \pm 6.2$ & $45.6 \pm 3.6$ & $87.4 \pm 2.5$ \\ 
			Age (Myr) & $6.2 \pm 0.6$ & $6.5 \pm 0.3$ & $5.0 \pm 0.2$ & $5.4 \pm 0.1$ & $5.7 \pm 0.3$ & $5.3 \pm 0.1$ & $4.6 \pm 0.0$ \\ 
			$\mathrm{12+\log(O/H)_S}$ & $7.28 \pm 0.12$ & $7.32 \pm 0.09$ & $7.32 \pm 0.12$ & $7.34 \pm 0.03$ & $-$ & $7.65 \pm 0.04$ & $7.57 \pm 0.02$ \\ 
			\hline
			\hline
			Region & $\mathrm{31}$ & $\mathrm{33}$ & $\mathrm{33}$ & $\mathrm{33}$ & $\mathrm{35}$ & $\mathrm{35}$ & $\mathrm{35}$ \\ 
			\hline
			Slit & IFU & PA198 & PA323 & IFU & PA211 & PA323 & IFU \\ 
			Pos.(arcsec) & -- & -16 -- -7 & 17 -- 19 & -- & -19 -- -10 & 20 -- 28 & -- \\ 
			E(B-V) & $0.00 \pm 0.02$ & $0.00 \pm 0.01$ & $0.17 \pm 0.02$ & $0.00 \pm 0.03$ & $0.01 \pm 0.02$ & $0.05 \pm 0.01$ & $0.00 \pm 0.02$ \\ 
			{[O \textsc{ii}] 3727,3729}/H$\beta$ & $1.72 \pm 0.22$ & $1.63 \pm 0.17$ & $-$ & $1.46 \pm 0.12$ & $-$ & $2.25 \pm 0.46$ & $1.59 \pm 0.10$ \\ 
			{[O \textsc{iii}] 5007}/H$\beta$ & $2.06 \pm 0.06$ & $2.39 \pm 0.04$ & $2.00 \pm 0.05$ & $3.05 \pm 0.10$ & $2.48 \pm 0.05$ & $1.75 \pm 0.02$ & $2.64 \pm 0.05$ \\ 
			{[N \textsc{ii}] 6583}/H$\alpha$ & $0.030 \pm 0.009$ & $0.027 \pm 0.002$ & $0.031 \pm 0.003$ & $0.021 \pm 0.007$ & $0.025 \pm 0.004$ & $0.028 \pm 0.002$ & $0.025 \pm 0.006$ \\ 
			{[S \textsc{ii}] 6717,6731}/H$\alpha$ & $0.131 \pm 0.007$ & $0.114 \pm 0.002$ & $0.121 \pm 0.004$ & $0.108 \pm 0.004$ & $0.094 \pm 0.004$ & $0.105 \pm 0.002$ & $0.108 \pm 0.003$ \\ 
			{[Ar \textsc{iii}] 7136}/H$\alpha$ & $-$ & $0.014 \pm 0.001$ & $0.015 \pm 0.002$ & $-$ & $0.020 \pm 0.004$ & $0.015 \pm 0.001$ & $-$ \\ 
			EW(H$\beta$) & $79.1 \pm 2.5$ & $108.5 \pm 2.5$ & $116.5 \pm 7.9$ & $170.6 \pm 7.1$ & $50.9 \pm 2.6$ & $147.3 \pm 7.6$ & $149.1 \pm 3.7$ \\ 
			Age (Myr) & $4.7 \pm 0.0$ & $4.3 \pm 0.0$ & $4.3 \pm 0.1$ & $3.8 \pm 0.1$ & $5.1 \pm 0.1$ & $4.0 \pm 0.1$ & $3.9 \pm 0.0$ \\ 
			$\mathrm{12+\log(O/H)_{Te}}$ & $-$ & $7.59 \pm 0.06$ & $-$ & $-$ & $-$ & $7.55 \pm 0.15$ & $7.50 \pm 0.05$ \\ 
			$\mathrm{12+\log(O/H)_S}$ & $7.68 \pm 0.08$ & $7.71 \pm 0.02$ & $7.69 \pm 0.03$ & $7.72 \pm 0.09$ & $7.69 \pm 0.05$ & $7.60 \pm 0.02$ & $7.72 \pm 0.06$ \\ 
			\hline
			\hline
			Region & $\mathrm{36}$ & $\mathrm{37}$ & $\mathrm{37}$ & $\mathrm{SB-3}$ & $\mathrm{SB-6}$ & $\mathrm{SB-6}$ &  \\ 
			\hline
			Slit & IFU & PA323 & IFU  & PA323 & PA211 & PA323 & \\ 
			Pos.(arcsec) & -- & 33 -- 39 & --  & 4 -- 9 & -30 -- -22 & 29 -- 31 & \\ 
			E(B-V) & $0.00 \pm 0.07$ & $0.03 \pm 0.02$ & $0.16 \pm 0.10$  & $0.12 \pm 0.08$ & $0.13 \pm 0.07$ & $0.00 \pm 0.06$ & \\ 
			{[O \textsc{ii}] 3727,3729}/H$\beta$ & $2.15 \pm 0.79$ & $-$ & $2.88 \pm 1.17$ & $-$ & $-$ & $-$ &  \\ 
			{[O \textsc{iii}] 5007}/H$\beta$ & $1.09 \pm 0.12$ & $0.68 \pm 0.03$ & $1.00 \pm 0.13$ & $0.39 \pm 0.09$ & $0.71 \pm 0.07$ & $0.96 \pm 0.08$ & \\ 
			{[N \textsc{ii}] 6583}/H$\alpha$ & $0.044 \pm 0.022$ & $0.044 \pm 0.005$ & $0.031 \pm 0.022$ & $0.042 \pm 0.013$ & $0.046 \pm 0.007$ & $0.056 \pm 0.011$ & \\ 
			{[S \textsc{ii}] 6717,6731}/H$\alpha$ & $0.295 \pm 0.051$ & $0.170 \pm 0.005$ & $0.202 \pm 0.045$ & $0.245 \pm 0.019$ & $0.194 \pm 0.011$ & $0.201 \pm 0.016$ &\\ 
			{[Ar \textsc{iii}] 7136}/H$\alpha$ & $-$ & $-$ & $-$ & $-$ & $-$  & $0.034 \pm 0.010$ & \\ 
			EW(H$\beta$) & $1.9 \pm 0.3$ & $144.4 \pm 24.5$ & $73.9 \pm 6.7$  & $12.6 \pm 2.8$ & $80.0 \pm 14.8$ & $88.9 \pm 19.2$ & \\ 
			Age (Myr) & $9.2 \pm 0.1$ & $4.0 \pm 0.3$ & $4.8 \pm 0.1$ & $7.2 \pm 0.3$ & $4.7 \pm 0.2$ & $4.6 \pm 0.2$ & \\ 
			$\mathrm{12+\log(O/H)_S}$ & $7.62 \pm 0.17$ & $7.41 \pm 0.04$ & $7.45 \pm 0.22$ & $7.22 \pm 0.12$ & $7.44 \pm 0.07$ & $7.62 \pm 0.08$ & \\ 
			\hline
		\end{tabular}
	\end{scriptsize}
\end{table*}

%
%
%

\subsubsection{\revone{Measurements of oxygen abundance}}

\revone{For measurements of $12+\mathrm{\log(O/H)}$ as indicator of the gas phase metallicity we utilize the S calibration from \cite{Pilyugin2016} -- an empirical method relying on the flux ratios of strong emission lines (\NII~6584\AA, \SII~6717,6731\AA, \OIII~5007\AA\, and H$\beta$). 
For two bright \HII regions (\#33 and \#35) we were able to also measure the faint auroral line \OIII~4363\AA\, and thus to measure the electron temperature $T_e$ and derive $12+\mathrm{\log(O/H)}_{Te}$ with the $T_e$-method following the relations in \cite{Izotov2006}.} 
Our measurements made using the S calibration show slightly higher values than those made using the $T_e$ method.  

Previous studies showed that DDO~53 has a low oxygen abundance $\mathrm{12 + \log(O/H)=7.82\pm0.09}$ \citep{Croxall2009}, however earlier estimates differ significantly showing $12 + \log\mathrm{(O/H)} = 7.52 - 8.59$ \citep{Hunter1985, Skillman1989, Hunter1999, Pustilnik2003, Saviane2008}. The median and standard deviation of oxygen abundance derived in our study from a larger sample of \HII regions $\mathrm{12 + \log(O/H)_S=7.59\pm0.16}$ is in good agreement with  $\mathrm{12 + \log(O/H)=7.52\pm0.08}$  obtained by \cite{Pustilnik2003} and with estimates made using the $T_e$ method for the brightest \HII regions, however it is lower than all other measurements in the literature. The distribution of the O/H measurements across different \HII regions show significant variations in the range of $\sim 7.3-7.8$~dex without any sign of a radial gradient. We find that our measured values of  $\mathrm{12 + \log(O/H)_S}$ strongly correlate with the surface brightness of corresponding \HII region (and anti-correlate with \SIIHa) and hence we expect that the faint nebulae might be biased by a significant DIG contribution. Given that, we also derive the flux-weighted mean value $\mathrm{12+\log(O/H)_S=7.67\pm0.04}$. This latter value is in a perfect agreement with its absolute magnitude $M_B$ according to the luminosity-metallicity reference relation \citep[e.g.][]{Berg2012} and still lower than measured by \cite{Croxall2009}.  Thus, we estimate the mean gas phase metallicity of DDO~53 to be $Z\sim 0.08-0.10 Z_\odot$, depending on whether we consider the probable contamination by DIG or not.

\revone{In Fig.~\ref{fig:BPT} we compare the photoionisation models from \cite{Gutkin2016} with the measured line flux ratios for individual \HII regions shown on the BPT diagnostic diagrams \citep*{BPT}.} 
We find that the model for $Z\sim0.13Z_\odot$ (orange line) could explain most of the data points, and there is a lack of agreement with the models for lower metallicity. Those points with \OIIIHb\, below the lower limit for the modelled curve have lower ionisation parameter than considered in the \cite{Gutkin2016} models, but could be fit by other models at the same metallicity providing we assume the low ionisation parameter values. Taking into account the well known problem of the discrepancy between oxygen abundance measurements made by empirical and model-based calibrations \cite[e.g.][]{Kewley2008}, we may conclude that our flux-weighted metallicity estimate is in agreement with the photoionisation models and that the nebulae with significantly lower values of $\mathrm{12+\log(O/H)}$ in Table~\ref{tab:ls_fluxes} are highly contaminated by DIG, and do not reflect real chemical inhomogeneity due to accretion of metal-poor gas or another process.

\subsubsection{\revone{Indication of shocks in the ISM}}

\begin{figure*}
	\includegraphics[width=\linewidth]{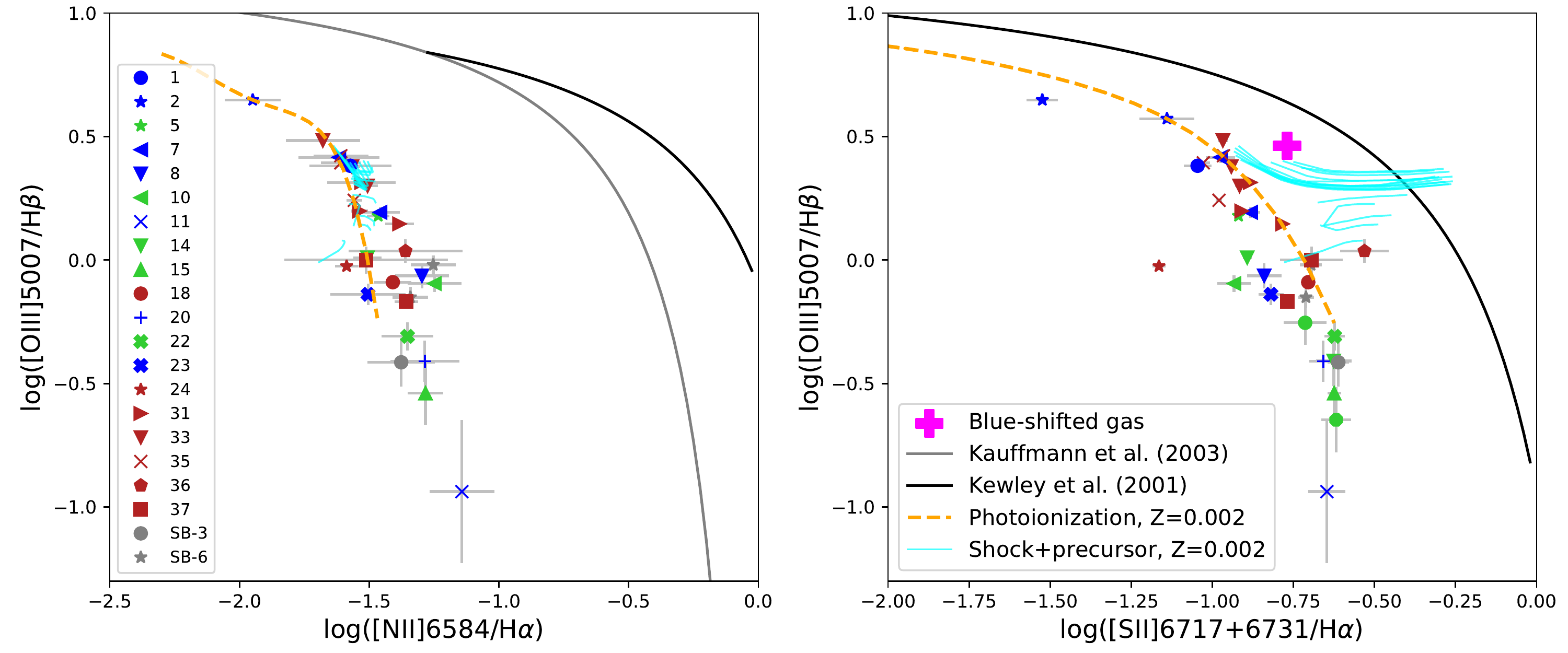}
	\caption{The diagnostic BPT diagram \citep*{BPT} \revone{together with its extension \citep{Veilleux1987}} showing the line ratios of \OIIIHb\, versus \NIIHa\, (left-hand panel) and \SIIHa\, (right-hand panel) for different \HII regions according to their numbers in Fig.~\ref{fig:regnames} and Table~\ref{tab:ls_fluxes}. Grey points for SB-3 and SB-6 correspond to the emission inside the identified superbubbles according to Fig.~\ref{fig:isigma}. The magenta cross on the right panel corresponds to the area of elevated \OIIIHb\, towards the region of the blue-shifted \HI and \Ha motions (see Fig.~\ref{fig:o3hb}). Black and grey curves represent the demarcation lines between photoionised regions (to the left of them) and those having high contribution of other  excitation mechanisms (for solar metallicity, taken from \citealt{Kewley2001} and \citealt{Kauffmann2003}, respectively). The orange dashed line is the polynomial approximation of the grid of photoionisation model from \citet{Gutkin2016} constructed for metallicity Z=0.002 ($12+\log\mathrm{(O/H)}\simeq7.8$), and the cyan lines are the model of shocks+precursor for the same metallicity and for shock velocities $V=100-500 \kms$ as given from \citet{Alarie2019}.}\label{fig:BPT}
\end{figure*}

\begin{figure}
	\includegraphics[width=\linewidth]{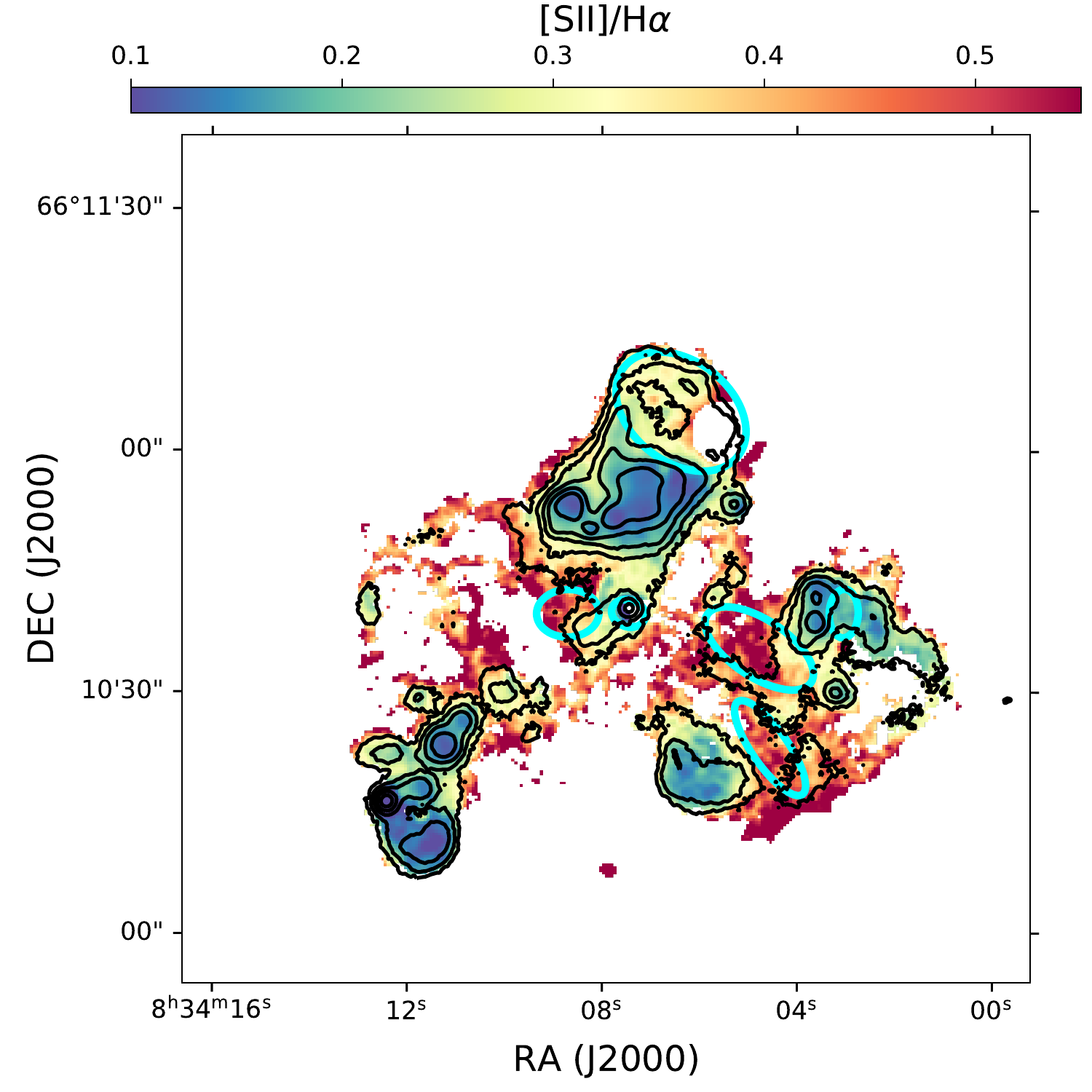}
	\caption{The flux ratio of \SIIHa\, as measured from narrow-band imaging observations. Black contours trace the levels of constant \Ha surface brightness. Cyan ellipses denote the borders of superbubbles as identified in Fig.~\ref{fig:isigma}.}\label{fig:s2ha}
\end{figure}

The estimated ages of the \HII regions demonstrate quite a uniform distribution around a value of $5.0\pm1.1$ Myr, with the older ages towards the low brightness area where we find signs of expanding superbubbles, and slightly younger ages for the brightest \HII regions. This is in agreement with the estimates made by \cite{Pustilnik2003}, yet we do not confirm their findings on the older age of the \HII regions towards the SE complex. Comparing the ages of the \HII regions and the estimates of the kinematic ages of the superbubbles (see Table~\ref{tab:superbubbles}), we may expect that the latter were developed when the \HII regions were relatively evolved and hence the supernovae explosions should have already occurred. Hence, the energy of supernovae explosions should contribute significantly to the energy balance of superbubbles, and we may expect to find signs of shock excitation towards them. 

The map of \SIIHa\, obtained from our narrow-band images (Fig.~\ref{fig:s2ha}) shows that all \HII regions have this flux ratio well below the typical value of 0.4 that is usually used for separation of supernovae remnants from photoionised nebulae \citep{Blair1981, Blair2004}. At the same time, the \SIIHa\, is slightly increased towards the identified superbubbles (the same also follows from the long-slit spectra for SB-3 and SB-6, see Table~\ref{tab:ls_fluxes}). This is especially evident for superbubble SB-2 where the \SIIHa\, reaches values above 0.5. In principle, the enhanced ratio of \SIIHa\, and other low-excitation ions is also usually observed in the DIG, probably related with the leaking ionising quanta from \HII regions (e.g., \citet{HidGam2007} and \citet{Haffner2009} specifically for DIG in DDO~53). However, since we also observe the presence of a blue-shifted broadened component in the \Ha profile in the centre of superbubble SB-2 (prof.~\#2 in Fig.~\ref{fig:FPI_profiles}), we  suggest there is a high contribution of shocks from supernovae there.

\revone{As is already mentioned, Fig.~\ref{fig:BPT} demonstrates the measurements of the line flux ratios made for the integrated spectra of the \HII regions as given in Table~\ref{tab:ls_fluxes} on the BPT diagrams, which are usually considered as the main diagnostic instrument allowing us to distinguish between different excitation mechanisms for ionised nebulae.} 
Different symbols correspond to different regions, and the grey symbols denoted with `SB' letters are related to the measurements made within the corresponding superbubbles. Those symbols sharing the same colour correspond to the same star-forming complex according to Fig.~\ref{fig:regnames} (red colour = complex N; green colour = SW; blue colour = SE). We do not find any significant difference in ionisation properties of the \HII regions within different complexes except that \OIIIHb\, (which could be considered as a proxy of hardness of the ionising radiation) is slightly lower for complex SW. As follows from the diagrams, all analysed nebulae in DDO~53 lie far below the demarcation lines from \cite{Kewley2001} and \cite{Kauffmann2003} within the photoionisation domain. However it is important to remember that the parametrization of these lines were obtained assuming solar metallicity, while DDO~53 is low-metallicity galaxy.

\revone{On the BPT diagram} we overlay (as cyan lines) the grid of shocks models at $Z\sim0.13Z_\odot$ that were computed by \cite{Alarie2019} for the same chemical abundance prescription as is in the \cite{Gutkin2016} models. Different lines correspond to different shock velocities. One may see that while in principle all the data points are well described by pure photoionisation, at least for bright regions in complex N the shocks acting in such a low-metallicity environment could produce the same line ratios, and hence it is impossible to distinguish between the contribution of these two mechanism without considering additional source of information (e.g., local gas kinematics as described in Sec.~\ref{sec:local_kin}). Since we detected the presence of a prominent \Ha line asymmetry and a blue-shifted cloud (Figs.~\ref{fig:asymm},\ref{fig:isigma}), as well as the underlying broad component between the bright \HII regions \#31, 33, 35 (prof.~\#4 in Fig.~\ref{fig:FPI_profiles}), we suggest that shocks contribute to the overall gas excitation in this area where the most intense  ongoing star formation occurs. The origin of these shocks could be related to either feedback from O stars, or from the interaction with the \HI tail.

\begin{figure*}
	\includegraphics[width=\linewidth]{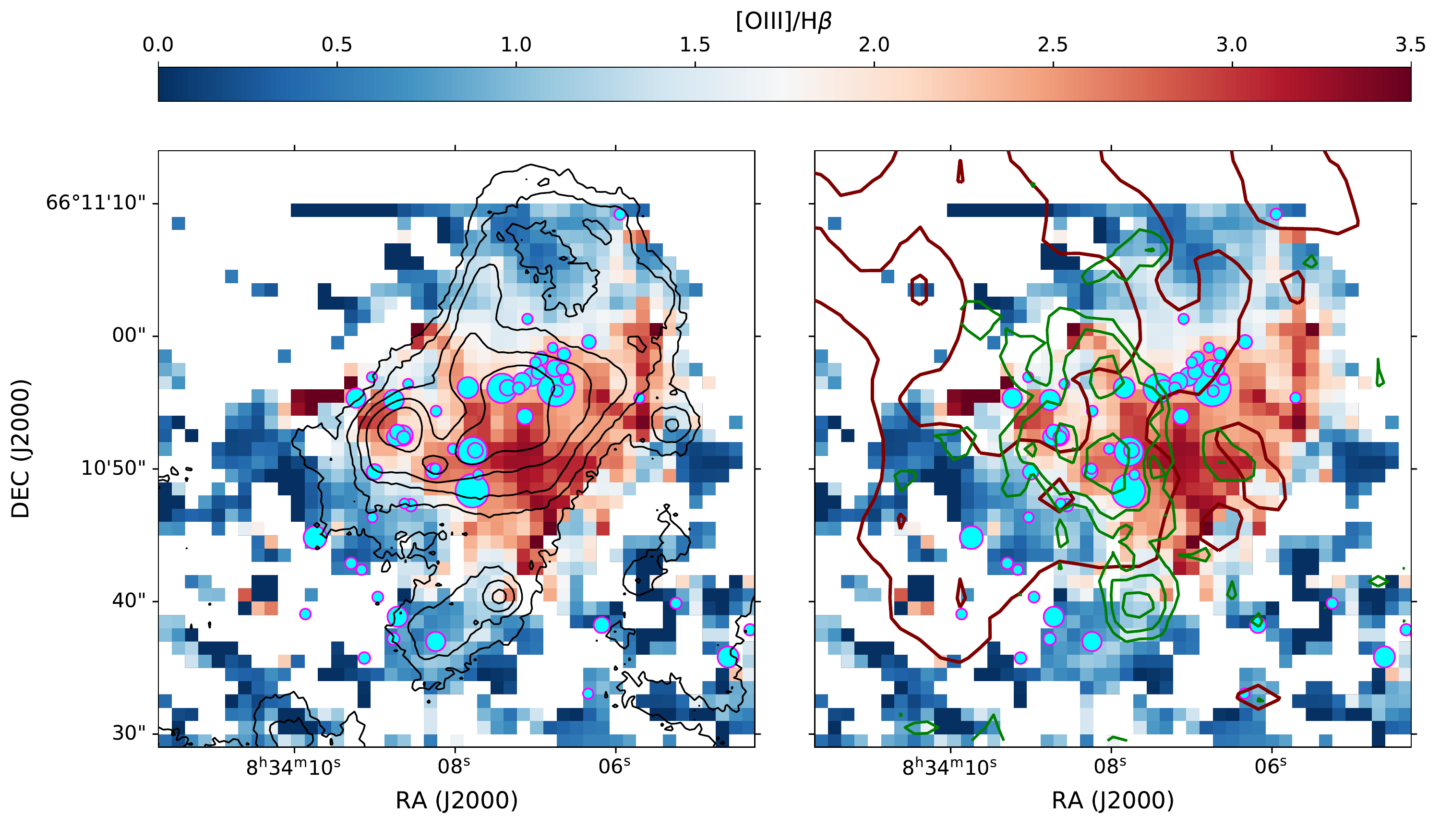}
	\caption{The distribution of the \OIIIHb\, lines flux ratio in the star-forming complex N as derived from IFU (PPAK/PMAS) data. \textbf{Left-hand panel}: black contours show the lines of constant \Ha surface brightness; cyan circles show the locations of identified O stars (size of the symbols correlate with their $M_V$). \textbf{Right-hand panel}: Brown contours show the lines of the constant \HI density of the \HI tail; green contours correspond to the asymmetry of the \Ha profile as given in Fig.~\ref{fig:asymm}. 
	}\label{fig:o3hb}
\end{figure*}

We also consider the ionisation structure of \revone{the brightest} complex N in more details. We do not see any significant enhancement of the \SIIHa\, flux ratio beyond the \HII regions (Fig.~\ref{fig:s2ha}), but the IFU data reveal elevated \OIIIHb\, at the edge of the bright nebulae rather than toward the central part (Fig.~\ref{fig:o3hb}). The highest \OIIIHb\, flux ratio in the galaxy (after the \HII region \#2) is observed here, exactly at the place where all three peculiarities in gas kinematics (the beginning of the \HI tail, the extended region of \Ha line profile asymmetries, and the area of blue-shifted non-circular motions in the ionised gas) were detected. Note that while the largest number of identified O stars concentrate in this complex, they are located outside the area of the highest \OIIIHb\, allowing us to suggest that these stars are unlikely to be the only sources for the observed gas excitation. Since shocks could also elevate \OIIIHb\, \citep{Allen2008}, we consider this scenario. The position of the most prominent extended region of enhanced \OIIIHb\, line ratio (coinciding with the area of the blue-shifted ionised gas) on the \OIIIHb\, vs \SIIHa\,  diagram is shown by magenta cross in Fig.~\ref{fig:BPT}. This region is shifted towards the top-right part of the diagram compared to the rest of the \HII regions in the complex and is better explained by the grid of shock models rather than by the models of pure photoionisation. From this and from the analysis of the gas kinematics we conclude that the shocks are observed at the south-western part of complex N.

\subsubsection{\revone{Peculiar \HII regions}}

\begin{figure}
	\includegraphics[width=\linewidth]{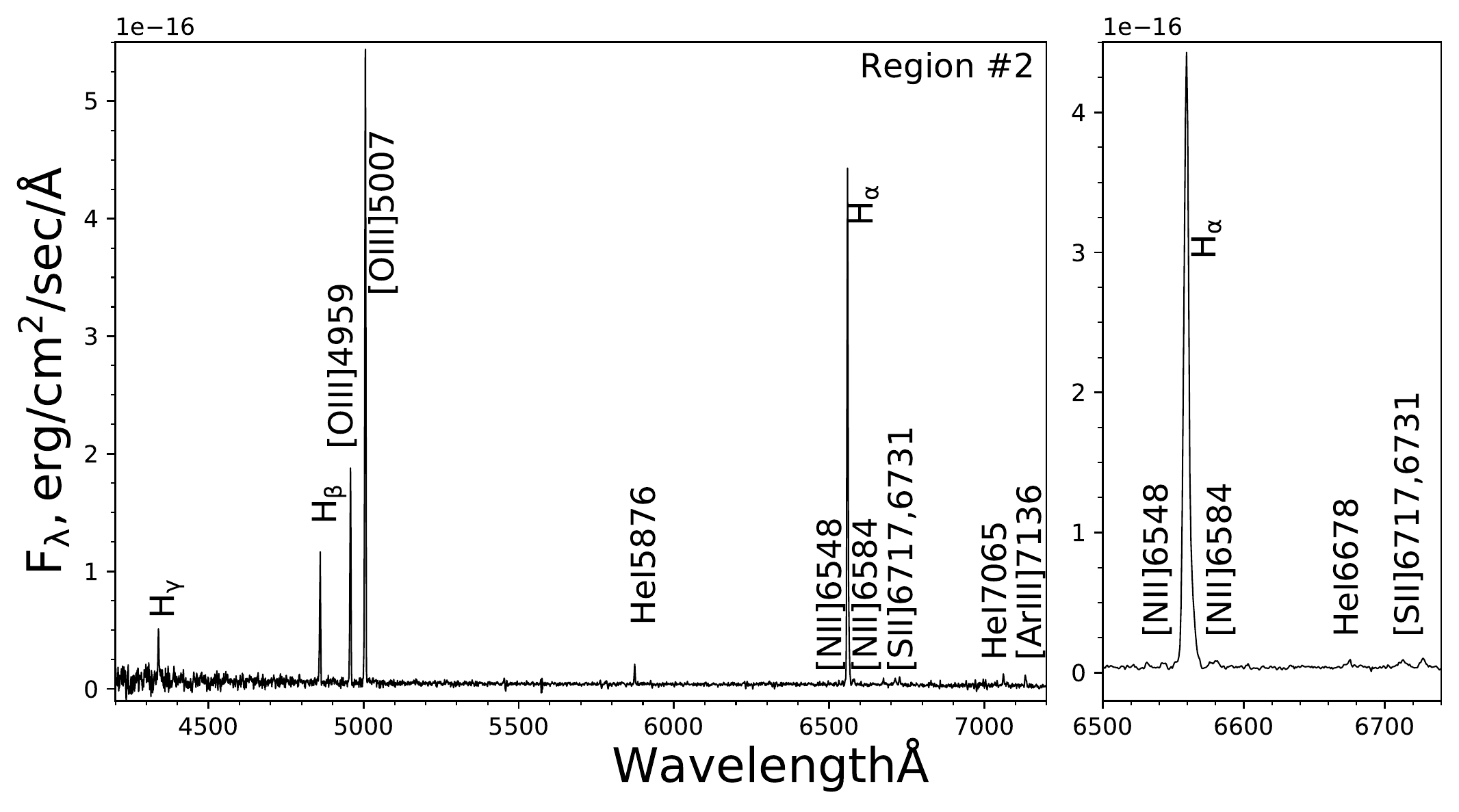}
	\includegraphics[width=\linewidth]{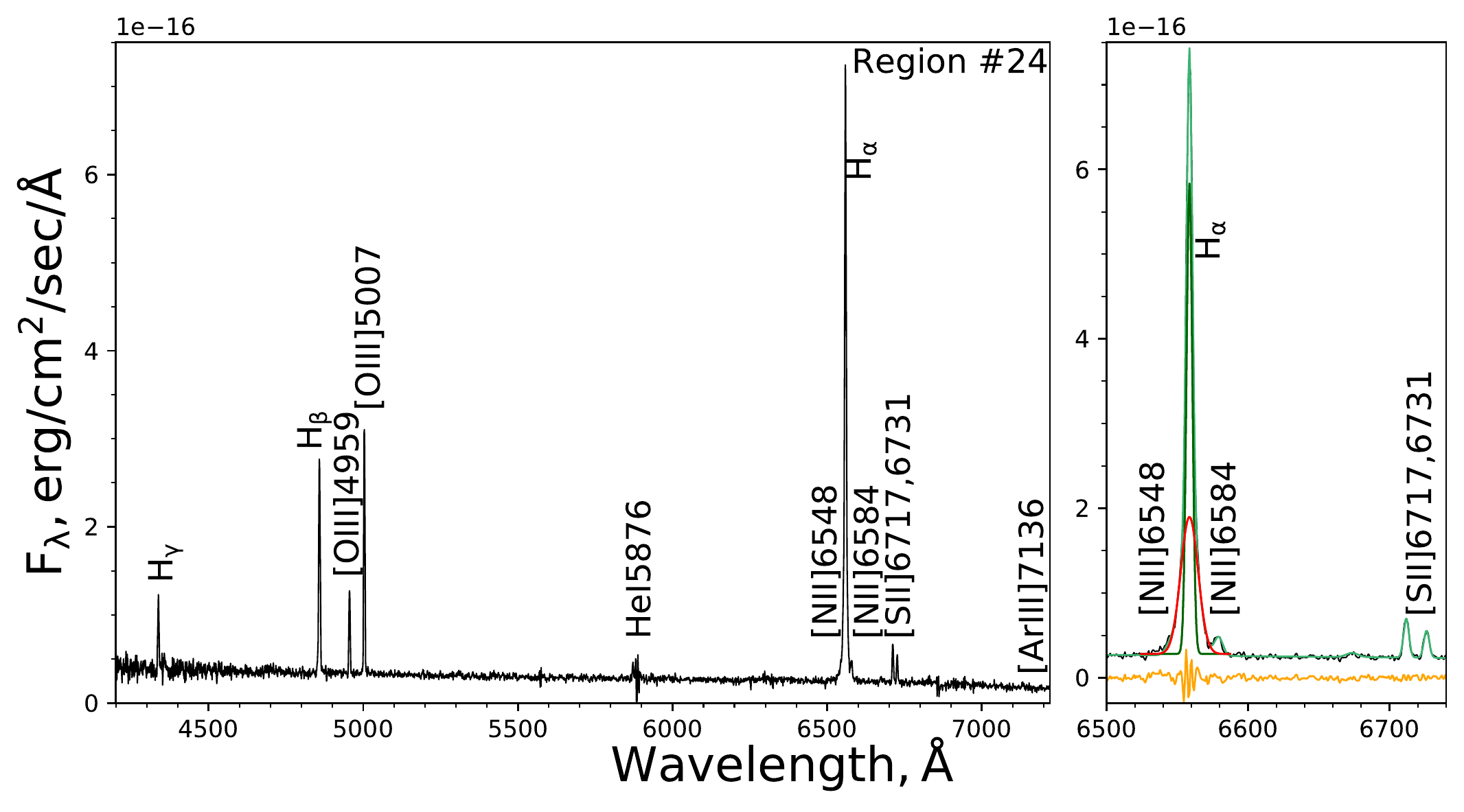}
	\caption{Example of the spectra for \HII regions \# 2 (top) and \#24 (bottom) extracted from the long-slit spectra. The \Ha line of the spectrum of region \#2 was fit with both narrow and broad Gaussians.
	}\label{fig:ls_stars}
\end{figure}

Finishing the analysis of the excitation state of the ionised gas, we describe two \HII regions having peculiarities in their spectra -- \#2 and \#24. Their spectra integrated over the area crossed by the slits PA=323 and PA=198, respectively, are given in Fig.~\ref{fig:ls_stars}.

\textit{Region \#2} demonstrate the highest \OIIIHb\, in the galaxy and deviates to the top-left from the rest \HII regions on BPT diagrams, yet it still could be well fit by the same photoionisation model. A local peak of $E(B-V)$ is also observed towards this region, however it is not detected by IFU observations (probably because of the different area of spectra extraction). Together with a high \OIIIHb, this region has the lowest \SIIHa\, and \NIIHa\, line ratios in DDO~53, thus suggesting that most of the gas there is in a high excitation state. From IFU observations we reveal high [O~\textsc{iii}]/[O~\textsc{ii}]$\sim18$. Such a large value is indicative that the region is density-bounded and hence should produce a significant amount of leaking LyC quanta \citep{Jaskot2013}.

\textit{Region \#24} was already mentioned in Sec.~\ref{sec:local_kin} as having the most prominent two-component \Ha line profile with the second component  broadened and red-shifted (prof.~\#3 in Fig.~\ref{fig:FPI_profiles}). This region coincides with the peak of the \Ha line asymmetry map (Fig.~\ref{fig:asymm}) and with one of the detected superbubbles (SB-5). The long-slit spectrum also clearly reveals a broadened component under the Balmer lines. On the BPT diagrams the \SIIHa\, ratio for the region stands out remarkably from the rest of \HII regions towards lower values, and thus points to probably having lower gas metallicities.  However the estimates given in Fig.~\ref{fig:BPT} and Tab.~\ref{tab:ls_fluxes} were obtained for integral fluxes in each involved lines, without decomposition onto the components. Assuming that the true flux ratio comes from the narrow components, we decompose the \Ha onto two Gaussian (as shown in Fig.~\ref{fig:ls_stars}) and estimate the resulting values of \SIIHa$_\mathrm{narrow}\simeq0.13$ and \NIIHa$_\mathrm{narrow}\simeq0.06$. With these values this region would occupy the same area on BPT diagrams as the \HII regions \#8 and \#10 and thus it should have higher oxygen abundance consistent with the photoionisation model. Given that the second broad component of the \Ha line is significantly red-shifted and extended well beyond the \HII region (as traced by \Ha asymmetry map), we may suggest that this underlying component is not related with the \HII region and the observed picture is a result of the projection along the line-of-sight of the regular \HII region and the faint region of non-circular motions.

\section{Discussion}\label{sec:discussion}

\subsection{A supergiant shell of ionised gas: large-scale outflow or the result of leaking LyC quanta?}\label{sec:feedback}

Our \Ha narrow-band images revealed faint ionised gas encircling the galaxy (Sec.~\ref{sec:HII}). This emission \revone{appears as} 
a supergiant shell of ionised gas. Its size (about 2 kpc) is significantly larger than for the biggest ionised supershell in the LMC \citep{Book2008},  and comparable with the one discovered in the galaxy Holmberg~II \citep{Egorov2017}. In this section we discuss the possible origin of such a giant emission structure.  

Through the influence of the winds and supernovae, massive stars create superbubbles of about 100~pc size in the ISM, while the merging of such structures could produce the kpc-sized supergiant shells observed in many nearby galaxies in both ionised and neutral gas.  Continuous ejection of momentum and energy leads to the development of large-scale outflows in the form of galactic fountains or winds. Comparing the optical and \HI maps for 12 dwarf galaxies, \cite{McQuinn2019} found that many of them demonstrate an excess of \Ha emission at their periphery -- this was interpreted as a signature of galactic fountains or winds (depending on the escape velocities of the gas). The discovered supershell of ionised gas in DDO~53 is also observed at the periphery of the galaxy and surrounds the stellar disc and bright \HI clouds, thus producing an excess of \Ha emission in the same way as in \cite{McQuinn2019}. 

Assuming that the detected ionised supershell in the galaxy DDO~53 is indeed an outflow, we estimate the mass-loss rate ($\dot{M}_{out}$) and mass-loading factor ($\eta=\dot{M}_{out}/SFR$) to check if the ongoing star formation is sufficient to drive such an outflow.  
With our angular resolution and S/N ratio in the \Ha line we cannot trace precisely the shape of the supershell. Since its brightest parts could be encircled by an ellipse having axis ratio and position angle in accordance with inclination and PA of the galaxy (see Sec.~\ref{sec:HII} and Fig.~\ref{fig:outflow}), we can assume a spherically symmetrical outflow in the form of a thin shell having radius of $R\sim1$~kpc and thickness of $h\sim0.2R$ that is typical for superbubbles \cite[e.g.][]{Churchwell2006, Krause2013} and consistent with our estimates derived from the smoothed \Ha map. We can estimate $\dot{M}_{out}$ from the total  mass of the outflow dividing it by the time required  for crossing the shell ($\tau=h/v_{out}$ where $v_{out}$ is the outflow velocity):
\begin{equation}
	\dot{M}_{out} = 4 \pi R^2 \mu m_H  <n_p> v_{out}, 
\end{equation}
where $m_H$ is the mass of hydrogen atoms, $\mu=1.4$ to account for the Helium contribution, $<n_p>$ is the average number density of ionised hydrogen atoms which has an unknown value. However, it could be derived from the surface brightness of the \Ha line $I(\mathrm{H\alpha})$ in the following way.

\begin{equation}
	I(\mathrm{H\alpha}) = \frac{h \nu_\mathrm{H\alpha}}{4 \pi} \int n_p n_e \alpha^{eff}_\mathrm{H\alpha} ds \simeq \frac{h \nu_\mathrm{H\alpha}\alpha^{eff}_\mathrm{H\alpha}L}{4 \pi} <n_e^2>,
\end{equation}
where $I(\mathrm{H\alpha})$ is expressed in erg~cm$^{-2}$~s$^{-1}$~sr$^{-1}$, the effective recombination coefficient  is \mbox{$\alpha^{eff}_\mathrm{H\alpha} \simeq 1.17\times10^{-13}$~cm$^3$~s$^{-1}$} for electron temperature $Te = 10000$~K \citep{Osterbrock2006}, and $L$ is the optical path toward the supershell rim where we measure the surface brightness. Taking into account the spherical thin-shell geometry, we can derive that $L=2h\sqrt{1+2R/h} \simeq 1.33R$ with our assumption of $h \sim 0.2R$. Due to clumping of the ISM, $<n_e^2> \ne <n_e>^2$, but they are related with each other through the filling factor. According to \cite{KadoFong2020}, $<n_e>\simeq 0.2 \sqrt{<n_e^2>}$ for the DIG. Hence, 
\begin{equation}
	<n_p> \simeq <n_e> \simeq 1.23\times 10^8 \sqrt{I(\mathrm{H\alpha})/R}, 
\end{equation}
where $I(\mathrm{H\alpha})$ is in erg~cm$^{-2}$~s$^{-1}$~arcsec$^{-2}$ and R is in pc.


Only the few brightest clumps in the supershell were detected in our FPI data making it impossible to measure the outflow velocity. We can only roughly estimate its upper limit as \mbox{$v_{out}\sim20 \kms$} assuming that the  earlier mentioned \HI `tail' (separated by this value from the bulk of the H~\textsc{i}) is related to the supershell (see the next section for a discussion on that), or that the maximum deviations of the map of residual velocities in \Ha line are related with such a large-scale outflow (see Fig.~\ref{fig:vel}). 

Combining the estimates of all involved parameters together, we derive  $\dot{M}_{out} \simeq 0.06 M_\odot$ and $\log(\eta) \simeq 1.05$, which is slightly higher than the values derived by \cite{McQuinn2019} for several other dwarf galaxies of similar circular velocity, but in a good agreement with the simulations by \cite{Christensen2016}. Note that the usage of a lower filling factor like in \cite{McQuinn2019} will bring our estimate into perfect agreement with their results.

Thus, we conclude that the 2 kpc sized \revone{supershell-like structure} of ionised gas could be a large-scale outflow driven by the ongoing star formation. \revone{However this scenario still might be challenged by comparison of the \Ha and \HI flux maps. As it follows from Fig.~\ref{fig:hi_cnahhels}, there is prominent \HI emission visible in blue-shifted channels at the periphery of DDO~53 beyond the \HI tail and the known \HI supershells. Some of these gas clouds correlate with peaks in the \Ha emission of the ionised supershell under discussion. Hence, we may expect that at least in part the \Ha emission could also be related to the ionisation of the surrounding atomic gas by leaking quanta. As we demonstrated in Sec.~\ref{sec:HII}, the amount of leaking ionising quanta from the star-forming complexes is enough to explain the emission of such diffuse structures.} In reality, both effects might contribute, and deeper \HI data and more accurate measurements of \Ha velocities are required to distinguish between them.
It is worthwhile to also note that the ionisation of the faint outer \HI gas in principle might also take place in other studies of some of the dwarf galaxies lacking the deep \HI data at their periphery, but having deep enough \Ha data. 

\subsection{Star formation history and its relation to the blue-shifted \HI tail}\label{sec:accretion} 

The kinematics and the localisation of the \HI tail allow one to suggest two explanations of its origin. It might be an external gas cloud falling onto the galaxy, or it might be resulted from a feedback-driven outflow due to the massive stars in complex N,
and thus it contains outflowing gas rather than inflowing gas. The blue-shifted velocities of the cloud are more consistent with the latter scenario, and the observed shocks could thus be related to the influence of the stellar winds and possible supernovae on the surrounding ISM, rather then with infalling gas. However it is hard to explain such a large extent and elongated shape in the \HI tail, which is more consistent with a gas accretion scenario.  Moreover, the observed SFR in the region N (see estimates below) is an order of magnitude less than it follows from eq. (1) in \cite{Veilleux2005} for  the mass-loss rate from the region resulting from feedback  $\dot{M_*}\sim0.26$ SFR.  Thus the observed star formation in complex N is insufficient to drive such an outflow, and the external origin of the \HI tail remains our favourable scenario.

DDO~53 is a rather isolated member of the M 81 group -- it has no known neighbouring galaxy within several tens of its optical radius \citep{Karachentsev2002}. Despite that, as we describe above, both the morphology and kinematics of its ISM are perturbed.  Since DDO~53 is located nearly between two subgroups of the M~81 group, \citet{Pustilnik2003} suggested that the current starburst there could be triggered by a tidal disturbance by the M~81 group as a whole, or by the interaction with the intergalactic medium (IGM), and these effects could be responsible for the observed peculiarities in the appearance of DDO~53. Estimating the density of the IGM at the periphery of the M~81 group to be too low, \cite{Begum2006} showed that ram pressure is unlikely to influence the gas in the galaxy. Instead they suggested that DDO~53 may be a product of a recent merger between two faint dwarf galaxies. From an analysis of the resolved \textit{HST} images \cite{Weisz2008} found 
that the star formation rate was enhanced about 1~Gyr ago. A dwarf-dwarf merger scenario can explain all the mentioned effects, including the presence of the blue-shifted \HI `tail' at the north part of the galaxy. However according to estimates by \cite{Weisz2008}, this event should have taken place not less than 1~Gyr ago. It is unclear whether the imprints of such interaction in the gaseous disc could survive for that time, as it is much longer than the dynamical timescale for the galaxy (about 120~Myr). 

Numerical studies shows that dwarf--dwarf mergers should not be rare, in particular for isolated galaxies, yet observational support of this process still remains very scarce -- only few low mass dwarf galaxies with evidences for  a recent major merger are known \citep[see, e.g.][]{Rich2012,Amorisco2014, Egorova2021}. \cite{Deason2014} found that $\sim 15-20$ per cent of `field' dwarf galaxies with $M_\star>10^6M_\odot$ in the Local Group likely experienced a major merger since $z\sim1$. However the authors note that it is unclear for how long major mergers would have significant residual impact on star formation, stellar kinematics, or morphology. According to simulations performed by \citet{Lotz2008} for more massive galaxies, the asymmetry of morphology could be observed at timescales of $0.3-1.1$~Gyr. A simulation of two gas-rich merging dwarf galaxies \citep{Bekki2008} also shows similar timescales and reproduces the morphology more or less consistently with that observed in DDO~53 (centrally concentrated regions of star formation surrounded by old stars and \HI envelope), but the expected SFR is much higher than what is observed in DDO~53. That author showed that such a scenario reproduces blue compact dwarf (BCD) galaxies.
In the case of DDO~53, an interaction in the past with a less massive companion (like it was established for DDO~68, see \citealt{Annibali2019}) seems to be a more favourable scenario. \cite{Starkenburg2016} showed in their models that the residuals of minor mergers for low mass galaxies can demonstrate a disturbed morphology in both gas and stars, and that this process leads to a modest enhancement of SFR that is comparable with the values observed in DDO~53. In their models, features like the \HI tail in DDO~53 could survive long enough, but the prominent stellar shells should also appear at the same time. They were not clearly detected in DDO~53, however the distribution of the red stars is asymmetrical \citep{Weisz2008}.

According to \cite{Weisz2008}, the SFR of DDO~53 was almost uniform and rather shallow during the last 1 Gyr, and the current burst of star formation occurred 25 Myr ago. While its morphological appearance could be explained by a minor merger that occurred $>1$~Gyr ago, the ongoing starburst is unlikely to be driven by this event. Alternatively, it could be triggered by the more recent accretion of a gas cloud from the IGM. It could be a remnant of the proposed merger event, or a gas cloud that resided in the galaxy vicinity. Despite the fact that DDO~53 is rather isolated galaxy, the later option doesn't appear to be unlikely. Thus, a perfect example of a starburst in an isolated galaxy driven by accretion of a gas cloud is IC~10 -- a nearby dwarf galaxy from the Local Group \citep{Ashley2014}. In such a scenario, the northern blue-shifted \HI tail might be directly related with the infalling gas cloud. Indeed, as it follows from Figs.~\ref{fig:hi_cnahhels}a, \ref{fig:HI_PV}, the tail is connected with the brightest star-forming complex N in the galaxy, accounting for about half of the current star formation in the galaxy, which allows us to speculate that neutral gas from this filament may provide the fuel for star formation.

We can roughly check whether accretion from the \HI tail is able to provide enough gas to sustain star formation in the brightest \HII complex. We assume a simple geometrical model, in which the `tail' is located in the galactic plane along the minor axis and has a cylindrical shape width $D=285$~pc (measured from blue-shifted channels in `derotated' robust-weighted \HI data). As it follows from \HI 21 cm line profile fitting and from the PV diagrams (Fig.~\ref{fig:hi_cnahhels}, \ref{fig:HI_PV}), the velocity of the `tail' is $V_\mathrm{LOS}\sim -20\ \kms$, with respect to the bulk \HI motions. This value can be transformed to the falling gas velocity in the galaxy plane $V_\mathrm{gas}=V_\mathrm{LOS}\times \cos(i)^{-1} \sim 25\ \kms$, assuming an inclination of $i=37^{\circ}$ (see Section~\ref{sec:HI}). Given the estimated \HI volume density $n_\mathrm{HI}\sim 0.25\ \mathrm{cm^{-3}}$ (assuming a constant scale height of the \HI disc, see Fig.~\ref{fig:vel}a), we estimate the gas infall rate as
\begin{equation}
	\dot{M}_\mathrm{gas} \simeq 1.43\times10^{-8} n_\mathrm{HI} D^2 V_\mathrm{gas}\ (M_\odot\ \mathrm{yr}^{-1}),
\end{equation}
which give us a value $\dot{M}_\mathrm{gas}\sim 0.007 M_\odot\ \mathrm{yr}^{-1}$. The value of $n_\mathrm{HI}$ might be underestimated if the size of the `tail' along the line of sight is significantly smaller than the adopted scale height of the \HI disc, which would lead to a higher value of $\dot{M}_\mathrm{gas}$.
Converting the \Ha line flux observed from the brightest complex of star formation to SFR we obtain
$\mathrm{SFR_{N}}\sim 0.002\ M_\odot/\mathrm{yr}$. Hence, comparing with the estimate of $\dot{M}_\mathrm{gas}$ we may conclude that such an infalling gas cloud will provide enough gas to sustain star formation at the observed rate. 

From the analysis of the small-scale ionised gas kinematics in complex N we find that the highly asymmetrical \Ha line profile, together with the area of the kinematically detached ionised gas motions, are observed towards the place where the \HI tail connects with the brightest \HII region (see Sec.~\ref{sec:local_kin}, Figs.~\ref{fig:asymm}, \ref{fig:isigma}). The spectral diagnostics reveal that the line ratios are consistent with gas excitation by shocks in that area (Sec.~\ref{sec:spectra_res}, Figs.~\ref{fig:BPT}, \ref{fig:o3hb}). These findings are consistent with a scenario where the \HI tail is a gas cloud falling onto the galactic disc and thus producing shock waves and triggering a burst of star formation in complex N. It is also interesting that almost all main sequence O stars in this region of the galaxy are located along the shape encircling this \HI cloud (see right panel of Fig.~\ref{fig:o3hb}). \revone{Note that despite the presence of oxygen abundance variations in our measurements (see Sec.~\ref{sec:spectra_res}), we find neither any strong indications of dilution by accreted low-metallicity gas, nor expelled metals driven by a large-scale outflow (Sec.~\ref{sec:feedback}) -- higher quality IFU data are required to clarify the small-scale oxygen abundance variations.}

\section{Summary}\label{sec:summary}

We present results from an observational analysis of the interplay between massive stars, the interstellar and intergalactic medium, and ongoing star formation in the nearby relatively isolated low-metallicity dwarf galaxy DDO~53. Our multiwavelength analysis is based on narrow-band optical imaging and long-slit and Fabry-Perot spectroscopy performed with the 6-m telescope BTA (SAO RAS), integral-field spectroscopy with PPAK/PMAS at 3.5-m telescope (Calar Alto) and archival data of JVLA (in \HI 21~cm line) and \HST observations. We obtain the following results: 

\begin{enumerate}
 \item Six expanding superbubbles of ionised gas were identified in the galaxy by the elevated velocity dispersion in the \Ha emission line. In general, their location is consistent with the presence of previously detected \HI supershells \citep{Pokhrel2020}. We demonstrate that the mechanical energy input from the winds of the observed O-stars within the superbubbles is enough to drive their expansion, but supernovae should also play a significant role, at least in part of the superbubbles, as follows from the diagnostic emission line ratios.
 \item We analysed the excitation state and measured the oxygen abundance in more than half of 37 detected \HII regions in the galaxy. Since the metallicity estimate derived from the empirical S calibration \citep{Pilyugin2016} correlates with the \Ha surface brightness and with the \SIIHa\, line ratio, we suggest that the measurements could be highly contaminated by the presence of DIG in the galaxy, which is also responsible for large variation of the estimated oxygen abundance. To get rid of this, we measure a \Ha flux-weighted mean value of $12+\mathrm{\log (O/H)}_S = 7.67\pm0.04$ that is consistent with the photoionisation and shocks model grids that could describe our observations. Without the weighting the measurements by the \Ha flux we get a mean value of $12+\mathrm{\log (O/H)}_{S} = 7.59\pm0.16$ that is in a good agreement with the estimates made using $T_e$ method for two brightest nebulae. Our measurements give a lower oxygen abundance than was adopted previously \citep{Croxall2009} and is more consistent with the lowest estimates obtained by \citet{Pustilnik2003}. 
 \item The brightest complex of star formation in DDO~53 demonstrates  excitation from shocks both in emission line ratios diagnostics and in its ionised and neutral gas kinematics. We suggest that the shocks originate from the interaction of the ISM in the galaxy with the anomalous \HI cloud that is detected there, rather than by the influence of the feedback from massive stars.
 \item We investigated the origin of the anomalous \HI gaseous tail to the north of the galaxy. From the analysis of its \HI kinematics and its relation to the kinematics and excitation of the ionised gas, we argue that it represents a gas cloud falling onto the galaxy. This gas accretion event might be a reason for the enhanced star formation in the galaxy $\sim 25$~Myr ago and it also could supply the fuel for sustaining star formation in the brightest \Ha complex of  DDO~53. The origin of the gas cloud still remains unknown, but it could be the remnant of a previous merger event that  occurred more than 1~Gyr ago, or it could be captured from the intergalactic medium since the M~81 group is gas rich. 
 \item We discover a faint giant 2-kpc sized ionised supershell surrounding the galaxy. It is likely that this structure represents a large-scale feedback-driven outflow. Under this assumption we estimate the mass-loading factor and find a value of $\log(\eta)\simeq 1.05$ that is slightly higher than the values obtained by \citet{McQuinn2019} for 4 other dwarf galaxies of the same circular velocity as for DDO~53, and in better agreement with the simulations by \cite{Christensen2016}. However we also \revone{suggest} that, at least in part, this extended \Ha emission could originate from the ionisation of the dispersed \HI at the periphery of DDO~53 by the leaking quanta from the star-forming regions. 
 \item We detected two small-size \HII regions exhibiting broad underlying Balmer lines and peculiarities in their spectra. One of them has a high line flux ratio of [O~\textsc{iii}]/[O~\textsc{ii}]$\sim 18$ and appears to be a density bounded nebula, which could produce significant amount of leaking ionising quanta. The other \HII region is probably a regular \HII region projected onto the extended region of diffuse gas excited by shocks.  
\end{enumerate}

As follows from the presented analysis, both stellar feedback and external gas accretion are important processes regulating the appearance of the ISM, its gas kinematics and the ongoing star formation in the DDO~53 galaxy. At the same time, the galaxy DDO~53 appears to be quite rare object where one can observe the interplay between the star formation activity and the ongoing gas accretion. A growing number of observational studies report the discoveries of the the anomalous \HI gas towards the bright \HII regions \cite[e.g.][]{Sancisi2008, Lelli2012}, or dynamical and excitation peculiarities of the ionised gas \revone{\cite[e.g.][]{SA2014, Silchenko2019, Egorova2019}}, as indications of the recent gas accretion fuelling the star formation activity. Our results suggest that in DDO~53 we observe all these effects simultaneously.

\section*{Acknowledgements}

The authors thank  A. Burenkov, D. Oparin and R. Uklein for their assistance in the SCORPIO-2 observations, and B. Groves and K. Sandstrom for their contribution in acquiring the PPAK IFU data.
The analysis of the gas kinematics and overall discussion presented in this work was supported by the Russian Science Foundation (projects no. 19-72-00149).
KK and OE gratefully acknowledge funding from the Deutsche Forschungsgemeinschaft (DFG, German Research Foundation) in the form of an Emmy Noether Research Group (grant number KR4598/2-1, PI Kreckel) for support of the gas excitation measurements.  Observations with the 6-m  telescope  of the Special Astrophysical Observatory of the Russian Academy of Sciences carried out with the financial support of the Ministry of Science and Higher Education of the Russian Federation (including agreement No05.619.21.0016, project ID RFMEFI61919X0016). The renovation of telescope equipment is currently provided within the national project `Science'.  
Based on observations collected at the Centro Astron\'{o}mico Hispano-Alem\'{a}n (CAHA) at Calar Alto, operated jointly by Junta de Andaluc\'{i}a and Consejo Superior de Investigaciones Cient\'{i}ficas (IAA-CSIC).
This research made use of Astropy (\url{http://www.astropy.org}) a community-developed core Python package for Astronomy \citep{astropy:2013, astropy:2018}, and of \textsc{astrodendro}, a Python package to compute dendrograms of Astronomical data (\url{http://www.dendrograms.org}). 
We acknowledge the usage of the Hyperleda data base (\url{http://leda.univ-lyon1.fr}).

\section*{DATA AVAILABILITY}
The data underlying this article will be shared on reasonable request to the corresponding author. The reduced FPI data are available in SIGMA-FPI data base\footnote{\url{http://sigma.sai.msu.ru}} (Egorov et al., in preparation).

\bibliographystyle{mnras}
\bibliography{DDO53}

\label{lastpage}

\end{document}